\documentclass[twocolumn]{jpsj3}
\usepackage{amsmath}
\usepackage{amssymb}
\usepackage{graphicx}
\usepackage{dcolumn}
\usepackage{bm}
\newcommand{\llangle}{\langle\kern -.23em \langle}
\newcommand{\rrangle}{\rangle\kern -.23em \rangle}
\renewcommand{\vec}[1]{\boldsymbol{#1}}
\renewcommand{\mod}{\,{\rm mod}\,}

\renewcommand{\vec}[1]{\boldsymbol{#1}}
\newcommand{\sgn}{\mbox{sgn}}

\title{
Belief Propagation for Error Correting Codes and Lossy Compression \\ Using Multilayer Perceptrons 
}

\author{
  \textsc{Kazushi MIMURA}$^1$,
  \thanks{E-mail address: \tt mimura@hiroshima-cu.ac.jp}
  \textsc{Florent COUSSEAU}$^2$
  and 
  \textsc{Masato OKADA}$^{2,3}$
}

\inst{
$^1$ Graduate School of Information Sciences, 
     Hiroshima City University, \\
     3--4--1 Ohtsuka-Higashi, Asaminami-ku, Hiroshima 731--3194, Japan \\ 
$^2$ Division of Transdisciplinary Sciences, 
     Graduate School of Frontier Sciences, \\
     The University of Tokyo, 
     5--1--5 Kashiwanoha, Kashiwa-shi, Chiba 277--8561, Japan \\
$^3$ RIKEN Brain Science Institute, 
     2--1 Hirosawa, Wako-shi, Saitama 351--0198, Japan
}

\abst{
The belief propagation (BP) based algorithm is investigated as a potential decoder 
for both of error correcting codes and lossy compression, 
which are based on non-monotonic tree-like multilayer perceptron encoders. 
We discuss that whether the BP can give practical algorithms or not in these schemes. 
The BP implementations in those kind of fully connected networks unfortunately shows strong limitation, 
while the theoretical results seems a bit promising. 
Instead, it reveals it might have a rich and complex structure of the solution space via the BP-based algorithms. 
}

\kword{
belief propagation, 
lossy compression, 
error correcting code, 
committee machine, 
parity machine, 
replica method, 
statistical mechanics 
}

\begin{document}
\maketitle

%%%%%%%%%%%%%%%%%%%%%%%%%%%%%%%%%%%%%%%%%%%%%%%%%%%%%%%%%%%%%%%%%%%%%%%%%%%%%%%%%%%%%%%%%%%%%
\section{Introduction}

\par
In today's society, information processing is part of our everyday life. 
As the pool of data available to us grows exponentially within the years, 
it is vital to be able to store, recover, and transmit those data in an efficient way. 
With the birth of information theory subsequently to the pioneering work of Shannon \cite{Shannon1948}, 
methods to efficiently process information start to become widely studied. 
\par
It has been shown that it is possible to ensure error free transmission using a non zero code rate 
up to a maximum value which cannot be exceeded without resulting in an inevitable loss of information. 
This upper bound is known as the Shannon bound. 
The design of efficient and practical codes is still one of the main topics of information theory. 
For example, the Sourlas's code \cite{Sourlas1989} asymptotically attains the Shannon bound, which is for channels with very small capacity. 
A interesting feature of Sourlas's paper is that it showed the possibility to use methods from statistical physics 
to investigate error correcting code schemes. 
Following this paper, the tools of statistical mechanics have been successfully 
applied in a wide range of problems of information theory in recent years. 
For instance in the field of error correcting codes itself 
\cite{Kabashima2000, Nishimori1999, Montanari2000, Cousseau2008bis}, 
as well as spreading codes \cite{Tanaka2001, Tanaka2005, Mimura2005, Mimura2006cdma, Mimura2007}. 
\par
On the other hand, lossy compression, which is the counterpart of lossless compression which seeks error free compression, 
has been also discussed \cite{Shannon1959}. 
Its task is to compress a given message allowing a certain amount of distortion 
between the original message and the reconstructed messages after compression. 
An efficient lossy compression scheme should be able to keep the compression rate 
as large as possible while keeping the distortion as small as possible. 
This is a typical trade-off optimization problem between the desired fidelity criterion and the compression rate. %[MODIFIED]
As in the reference \cite{Shannon1948}, 
Shannon derived an upper bound which gives the optimal achievable compression rate 
for a fixed distortion, i.e., a fixed fidelity criterion. 
Recently, statistical mechanical techniques were applied to these kind of problems 
with interesting results 
\cite{Murayama2003,Murayama2004,Hosaka2002,Hosaka2005,Mimura2006,Cousseau2008}. 
\par
This paper focuses on error correcting code and lossy compression 
where non-monotonic tree-like committee machines or parity machines are used 
as encoder and decoder respectively (for a thorough review on these kind of neural networks, see the reference \cite{Engel2001}). 
It has been analytically shown that in both error correcting code 
and lossy compression cases, this kind of schemes can reach the Shannon bound 
under some specific conditions \cite{Cousseau2008bis,Cousseau2008}. 
While these results are interesting from a theoretical point of view, 
the complexity of a formal encoder/decoder prevents these schemes from being practical. 
A formal way of encoding/decoding information would require an amount of time 
which grows exponentially with the size of the original message. 
One possible solution is to use the popular belief propagation (BP) algorithm 
in order to approximate the marginalized posterior probabilities of the appropriate Boltzmann factor 
which describes the behavior of the scheme.
\par
The BP algorithm is proved to be exact and is guaranteed to converge 
only for probability distribution which can be represented into a factor graph with no loop, i.e., a tree. 
This is not the case for schemes based on the above kind of neural networks 
as they are densely connected and necessarily contains loops, i.e., their corresponding factor graph is not a tree. 
\par
Nonetheless, the BP is known to give excellent approximating performance 
in the case of sparsely connected graph and have been successfully applied in decoding 
low density parity check codes (LDPC) for example. 
On the other hand, it is known that the approximation given by the BP 
in the case of more densely connected graphs is sometimes more mitigated. 
Several problems of sub-optimal solutions or simply convergence failures arises. 
\par
However, despite those issues, 
it is considered that investigating the BP algorithm on such kind of densely connected schemes 
is still interesting from a statistical physical point of view 
and provides precious insight into the solution space structure of such kind of systems. 
So far, only the BP-based encoders of lossy compression, based on 
both of the low-density generator-matrix (LDGM) code \cite{Murayama2003} and the simple perceptron \cite{Hosaka2006}, 
have been discussed. %[ADDED]
Both BP-based encoders for lossy compression based on the multilayer perceptron (MLP) and 
BP-based decoders for error correcting codes based on the MLP have never investigated yet. %[ADDED]
In this paper, we discuss that whether the BP can give practical algorithms or not 
in both error correcting codes and lossy compression based on the MLP. %[ADDED]
\par
The paper is organized as follows. 
Section \ref{sec:structure} introduces non-monotonic tree-like multilayer perceptron networks used throughout the paper. 
Section \ref{sec:ECCandLC} exposes the frameworks of error correcting code and lossy compression. 
Section \ref{sec:BP} introduces the belief propagation algorithm 
and section \ref{sec:performance} states the results obtained by the algorithm in both schemes. 
Section \ref{sec:discussion} is devoted to discussion and conclusion.

%%%%%%%%%%%%%%%%%%%%%%%%%%%%%%%%%%%%%%%%%%%%%%%%%%%%%%%%%%%%%%%%%%%%%%%%%%%%%%%%%%%%%%%%%%%%%
\section{Structure of multilayer perceptrons}
\label{sec:structure}

\par
In this section we introduce the kind of network we will use throughout the paper. 
Tree like perceptrons were already studied thoroughly by the machine learning community over the years. 
It is known that a feed-forward network with a single hidden layer made of sufficiently many units 
is able to implement any Boolean function between input layer and output. 
\par
The choice to use perceptron like networks for problem of information theory was already proposed by Hosaka et al. \cite{Hosaka2002}. 
They used a simple perceptron to investigate a lossy compression scheme. 
One of the most interesting feature of their work was the use of the following non-monotonic transfer function for the perceptron, 
\begin{equation}
  f_k (x) =
  \left\{
  \begin{array}{rl}
     1, & \quad |x| \leq k \\
    -1, & \quad |x| > k
  \end{array}
  \right.
  \label{f_k}
\end{equation}
where $k$ is a real parameter, controlling the bias of the output sequence. 
This choice of a non-monotonic transfer function was inspired by previous well known results 
within the machine learning community such as the improved storage capacity achieved by non-monotonic networks. 
They choose this modified version of a reversed wedge perceptron (see \cite{Engel2001} for a description of those networks) for several reasons. 
The first one was motivated by the need to be able to control the bias of the output sequence easily 
(which is achieved by tuning the parameter $k$). 
The second reason was motivated by the claim that a zero Edwards-Anderson (EA) order parameter is needed, 
thus reflecting optimal compression within the codeword space (meaning that codewords are uncorrelated in the codeword space). 
The use of (\ref{f_k}) ensures mirror symmetry ($f_k(x) = f_k(-x)$) and is likely to give rise to a zero EA order parameter (see \cite{Hosaka2002}).
\par
\begin{figure}[t]
  \vspace{0mm}%<--space
  \begin{center}
  \includegraphics[width=0.5\linewidth,keepaspectratio]{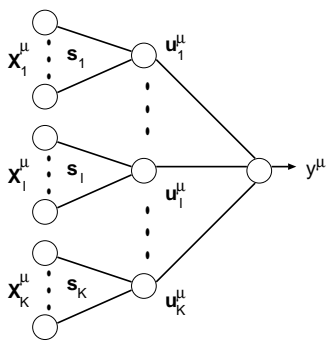}
  \end{center}
  \caption{General architecture of the treelike multilayer perceptron with $N$ input units
  and $K$ hidden units.}
  \label{fig:net}
  \vspace{0mm}%<--space
\end{figure}

Subsequently, non-monotonic tree-like perceptrons were successfully used 
in a lossy compression scheme and error correcting code scheme 
using the same kind of non-monotonic transfer function \cite{Cousseau2008, Cousseau2008bis}. 
This paper uses the same networks, which are all derived from the general architecture given by Figure \ref{fig:net}. 
\par %[MOVED]
In each of these networks, the coupling vector $\vec{s}$ is split into $\vec{s}=(\vec{s}_1,\ldots,\vec{s}_l,\ldots,\vec{s}_K)$ 
where each $\vec{s}_l=(s^1_l,\ldots,s^i_l,\ldots,s^{N/K}_l)$ is 
a $N/K$-dimensional binary vector of Ising variables (i.e.: $\pm 1$ elements). 
In the same way, the input vector $\vec{x}^{\mu}=(\vec{x}_1^{\mu},\ldots,\vec{x}_l^{\mu},\ldots,\vec{x}_K^{\mu})$, $\mu \in \{1,\cdots,M\}$ 
is also made of $N/K$-dimensional binary vector 
$\vec{x}_l^{\mu}=(x^{\mu}_{1 l},\ldots,x^{\mu}_{i l},\ldots,x^{\mu}_{N/K,l})$ of Ising variables. 
The output of the network is then given by the scalar $y^{\mu}$ which is also $\pm 1$. 
The $\sgn$ function denotes the sign function taking $1$ for $x \geq 0$ and $-1$ for $x <0$. 
We use the Ising expression (bipolar expression) $\{1,-1,\times \}$ instead of the Boolean expression $\{0,1,+\,(\mod 2)\}$ to simplify calculation. 
Consequently, the Boolean $0$ is mapped onto $1$ in the Ising framework while the Boolean $1$ is mapped to $-1$. 
This mapping can be used without any loss of generality. 
%From this architecture, %[REMOVED]
We investigate three different networks which are given by the followings: 
\par
(I) Multilayer parity tree with non-monotonic hidden units (PTH).
\begin{equation}
  y^{\mu} (\vec{s}) \equiv
  \prod_{l=1}^K f_k \left( \sqrt{\frac {K} {N}} \, \vec{s}_l \cdot \vec{x}_l^{\mu} \right) .
  \label{parity}
\end{equation}
\par
(II) Multilayer committee tree with non-monotonic hidden units (CTH).
\begin{equation}
  y^{\mu} (\vec{s}) \equiv
  \sgn \left( \sum_{l=1}^K f_k \left[ \sqrt{\frac {K} {N}} \, \vec{s}_l \cdot \vec{x}_l^{\mu} \right] \right) .
  \label{committeeH}
\end{equation}
Note that in this case, if the number of hidden units $K$ is even,
then there is a possibility to get $0$ for the argument of the sign
function. We avoid this uncertainty by considering only an odd
number of hidden units for the committee tree with non-monotonic
hidden units in the sequel.
\par
(III) Multilayer committee tree with a non-monotonic output unit (CTO).
\begin{equation}
  y^{\mu} (\vec{s}) \equiv
  f_k \left( \sqrt {\frac 1 K}\sum_{l=1}^K
  \sgn \left[ \sqrt{\frac {K} {N}} \, \vec{s}_l \cdot \vec{x}_l^{\mu}
  \right] \right).
  \label{committeeO}
\end{equation}

%%%%%%%%%%%%%%%%%%%%%%%%%%%%%%%%%%%%%%%%%%%%%%%%%%%%%%%%%%%%%%%%%%%%%%%%%%%%%%%%%%%%%%%%%%%%%
\section{Frameworks}
\label{sec:ECCandLC}

%%%%%%%%%%%%%%%%%%%%%%%%%%%%%%%%%%%%%%%%%%%%%%%%%%%%%%%%%%%%%%%%%%%%%%%%%%%%%%%%%%%%%%%%%%%%%
\subsection{Error correcting codes using multilayer perceptrons}

\par
In this section we show how non-monotonic tree-like perceptron can be used in an error correcting code scheme. 
\par
In a general scheme, an original message $\vec{s}^0 \in \{ -1,1 \}^N$ of size $N$ is encoded 
into a codeword $\vec{y}_0 \in \{ -1,1 \}^M$ of size $M$ by some encoding device. 
The aim of this stage is too add redundancy into the original data. 
Therefore, we necessarily have $M>N$. 
Based on this redundancy, a proper decoder device 
should be able to recover the original data 
even if it were corrupted by some noise in the transmission channel. 
The quantity $R=N/M$ is called the code rate and evaluates the trade-off between redundancy and codeword size. 
The codeword $\vec{y}_0$ is then fed into a channel where the bits are subject to some noise. 
The received corrupted message $\vec{y} \in \{ -1,1 \}^M$ (which is also $M$ dimensional) 
is then decoded using its redundancy to infer the original $N$ dimensional message $\vec{s}^0$. 
In other words, in a Bayesian framework, one try to maximize the following posterior probability, 
\begin{eqnarray}
  P(\vec{s}|\vec{y}) \propto P(\vec{y}|\vec{s}) P(\vec{s}). 
\end{eqnarray}
As data transmission is costly, 
generally one wants to be able to ensure error free transmission 
while transmitting the less possible bits. 
In other words, one wants to ensure error free transmission 
keeping the code rate as close as possible to the Shannon bound. 
\par
\begin{figure}[t]
  \vspace{0mm}%<--space
  \begin{center}
  \includegraphics[width=0.5\linewidth,keepaspectratio]{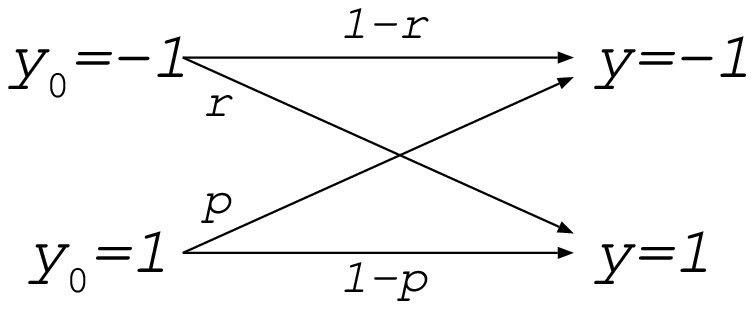}
  \end{center}
  \caption{The Binary Asymmetric Channel (BAC)}
  \label{fig:BAC}
  \vspace{0mm}%<--space
\end{figure}

In this paper we assume that the original message $\vec{s}^0$ 
is uniformly distributed on $\{ -1,1 \}^N$ 
and that all the bits are independently generated so that we have 
\begin{eqnarray}
  P(\vec{s}^0)= \frac1 {2^N}.
\end{eqnarray}
The channel considered is the Binary Asymmetric Channel (BAC) 
where each bit is flipped independently of the others with asymmetric probabilities. 
If the original bit fed into the channel is $1$, 
then it is flipped with probability $r$. 
Conversely, if the original bit is $-1$, it is flipped with probability $p$. 
Figure \ref{fig:BAC} shows the BAC properties. 
The binary symmetric channel (BSC) corresponds to the particular case where $r=p$. 
\par
\begin{figure}[t]
  \vspace{0mm}%<--space
  \begin{center}
  \includegraphics[width=0.8\linewidth,keepaspectratio]{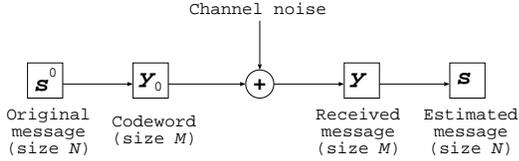}
  \end{center}
  \caption{Layout of the error correcting code scheme}
  \label{fig:ECCscheme}
  \vspace{0mm}%<--space
\end{figure}

 
Finally, the corrupted message $\vec{y}$ is received at the output of the channel. 
The goal is then to find back $\vec{s}^0$ using $\vec{y}$. 
The state of the estimated message is denoted by the vector $\vec{s}$. 
The general schematic outline of the scheme is shown in Figure \ref{fig:ECCscheme}.
From Figure \ref{fig:BAC} we can easily derived the following conditional probability,
$P(y^{\mu}|y^{\mu}_0) = \frac 1 2 + \frac {y^{\mu}} 2 [(1-r-p) y^{\mu}_0 + (r-p)]$, 
where we make use of the notations $\vec{y}_0 = (y_0^1,\ldots,y_0^{\mu},\ldots,y_0^M)$, 
$\vec{y} = (y^1,\ldots,y^{\mu},\ldots,y^M)$. 
Since we assume that the bits are flipped independently, 
we deduce $P(\vec{y}|\vec{y}_0) = \prod_{\mu=1}^M P(y^{\mu}|y^{\mu}_0)$. 
To encode the original message $\vec{s}^0$ into a codeword $\vec{y}_0$, 
we make use of the non-monotonic tree-like parity machine or committee machine neural networks already introduced. 
We prepare a set of $M$ input vectors $(\vec{x}^1,\ldots,\vec{x}^{\mu},\ldots,\vec{x}^M)$ 
which are drawn independently and uniformly on $\{ -1,1 \}$. 
This will play the role of the codebook. 
The original message $\vec{s}^0$ is used as the coupling vector of the network. 
Then, each input vector $\vec{x}^{\mu}$ is fed sequentially 
into the network generating a corresponding scalar $y_0^{\mu}$ 
at the output of the network finally resulting in a $M$-dimensional vector $\vec{y}_0$. 
This gives us the codeword to feed into the channel.
\par
The use of random input vectors is known to maximize 
the storage capacity of perceptron's network and since each $y^{\mu}_0$ is computed 
using the whole set of original bits $\vec{s}^0$, 
redundancy is added into the codeword. 
This makes such kind of scheme promising for error correcting task. 
A formal decoder should be able to decode the received corrupted message $\vec{y}$ 
by maximizing the posterior probability $p(\vec{s}|\vec{y})$, that is 
\begin{equation}
  \hat{\vec{s}} \equiv \mathop{\rm argmax}_{\vec{s} \in \{-1,1\}^N} \, p(\vec{s}|\vec{y}). \label{decoder}
\end{equation}
To keep notation as general as possible, 
as long as explicit use of the encoder is not necessary in computations, 
we will denote the transformation perform on the vector $\vec{s}$ 
by the respective tree-like perceptrons using the notation 
$\mathcal{F}_k ( \sqrt{\frac K N} \vec{s}_l \cdot \vec{x}_l^{\mu} )$. 
Here $\mathcal{F}_k$ takes a different expression for the three different types of network 
and this notation means all encoders depends on a real threshold parameter $k$. 
\par 
Since the relation between an arbitrary message $\vec{s}$ 
and the codeword fed into the channel is deterministic, 
for any $\vec{s}$, we can write 
$P(\vec{y}|\vec{s}) 
  = \prod_{\mu=1}^M \{ \frac 1 2 + \frac {y^{\mu}} 2 [(1-r-p) 
  \mathcal{F}_k ( \sqrt{\frac K N} \vec{s}_l \cdot \vec{x}_l^{\mu} ) + (r-p)] \}$. 
We finally get the explicit expression of the joint probability of the model as 
\begin{eqnarray}
  & & P(\vec{y},\vec{s}) 
      = \frac 1 {2^N} \prod_{\mu=1}^M \bigg\{ \frac 1 2 + \frac {y^{\mu}} 2 [(1-r-p) \nonumber \\
  & & \qquad \qquad \times \mathcal{F}_k \bigg( \sqrt{\frac K N} \vec{s}_l \cdot \vec{x}_l^{\mu} \bigg) + (r-p)] \bigg\} .
\end{eqnarray}
\par
The typical performance of this scheme was already studied 
using the Replica Method (RM) \cite{Cousseau2008bis} 
and it was shown that each of the three proposed network %[MODIFIED]
can reach the optimal Shannon bound at the infinite codeword length limit 
(when $N \to \infty$ and $M \to \infty$ while the code rate $R$ is kept finite) 
under some specific condition. 
\par
The PTH and the CTH were shown to reach the Shannon bound 
for any number of hidden units $K$ (any odd number of hidden units 
in the case of the CTH) if the threshold parameter $k$ 
of the non-monotonic transfer function is properly tuned. 
The CTO was shown to reach the Shannon bound 
when its number of hidden units $K$ becomes infinite 
and with a properly tuned threshold parameter $k$ only.

%%%%%%%%%%%%%%%%%%%%%%%%%%%%%%%%%%%%%%%%%%%%%%%%%%%%%%%%%%%%%%%%%%%%%%%%%%%%%%%%%%%%%%%%%%%%%
\subsection{Lossy compression using multilayer perceptrons}

\par
In this section we introduce the framework of lossy data compression \cite{Cover1991} 
and how non-monotonic tree-like perceptrons can be used for this purpose.
\par
\begin{figure}[t]
  \vspace{0mm}%<--space
  \begin{center}
  \includegraphics[width=0.85\linewidth,keepaspectratio]{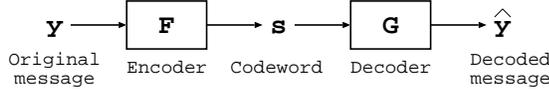}
  \end{center}
  \caption{Rate distortion encoder and decoder.}
  \label{fig:LossyScheme}
  \vspace{0mm}%<--space
\end{figure}

 
Let $\vec{y}$ be a discrete random variable defined on a source alphabet $\mathcal{Y}$. 
An original source message is composed of $M$ random variables, 
$\vec{y} = (y^1,\ldots,y^M) \in \mathcal{Y}^M$, and compressed into a shorter expression. 
The encoder compresses the original message $\vec{y}$ into a codeword $\vec{s}$, 
using the transformation $\vec{s}=\mathcal{F}(\vec{y}) \in \mathcal{S}^N$, where $N<M$. 
The decoder maps this codeword $\vec{s}$ onto the decoded message $\hat{\vec{y}}$, 
using the transformation $\hat{\vec{y}}=\mathcal{G}(\vec{s}) \in \hat{\mathcal{Y}}^M$. 
The encoding/decoding scheme can be represented as in Figure \ref{fig:LossyScheme}. 
In this case, the code rate is defined by $R=N/M$. 
A distortion function $d$ is defined as a mapping 
$d:\mathcal{Y} \times \hat{\mathcal{Y}}\rightarrow \mathbb{R}^+$. 
For each possible pair of $(y,\hat{y})$, it associates a positive real number. 
In most of the cases, the reproduction alphabet $\hat{\mathcal{Y}}$ 
is the same as the alphabet $\mathcal{Y}$ on which the original message $\vec{y}$ is defined.
\par
Hereafter, we set $\hat{\mathcal{Y}}=\mathcal{Y}$, and we use the Hamming distortion 
as the distortion function of the scheme. This distortion function is given by 
\begin{equation}
  d(y,\hat{y})=
  \left\{
  \begin{array}{rl}
    0, & \quad y=\hat{y}, \\
    1, & \quad y \neq \hat{y}, \\
  \end{array}
  \right.
\end{equation}
so that the quantity $d(\vec{y},\hat{\vec{y}}) = \sum_{\mu=1}^M d(y^{\mu},\hat{y^{\mu}})$ 
measures how far the decoded message $\hat{\vec{y}}$ is from 
the original message $\vec{y}$. 
In other words, it records the error made on the original message 
during the encoding/decoding process. The probability of error distortion 
can be written $E[d(y,\hat{y})]=P[y \neq \hat{y}]$ where $E$ represents 
the expectation. Therefore, the distortion associated with the code is defined 
as $D=E[\frac 1 M d(\vec{y},\hat{\vec{y}})]$, where the expectation is taken with respect to 
the probability distribution $P[\vec{y},\hat{\vec{y}}]$. $D$ corresponds to the 
average error per variable $\hat{y}^{\mu}$. Now we defined a rate distortion pair $(R,D)$ 
and we said that this pair is achievable if there exist a coding/decoding scheme 
such that when $M \to \infty$ and $N \to \infty$ (note that the rate $R$ is kept finite), 
we have $E[\frac 1 M d(\vec{y},\hat{\vec{y}})] \leq D$. 
In other words, a rate distortion pair $(R,D)$ is said to be achievable 
if there exist a pair $(\mathcal{F},\mathcal{G})$ such that 
$E[\frac 1 M d(\vec{y},\hat{\vec{y}})] \leq D$ in the limit 
$M \to \infty$ and $N \to \infty$.
\par
The optimal compression performance that can be obtained 
in the framework of lossy compression is given by the so-called 
\textit{rate distortion function} $R(D)$ which gives the best achievable code rate $R$ 
as a function of $D$ (Shannon bound for lossy compression). 
However, despite the fact that the best achievable performance is known, 
as in the error correcting code case, no clues are given 
about how to construct such an optimal compression scheme. 
\par
In this paper we assume that the original message $\vec{y}=(y^1,\ldots,y^{\mu},\ldots,y^M)$ 
is generated independently by an identically biased binary source, 
so that we can easily write the corresponding probability distribution, 
\begin{equation}
  P[y^{\mu}] = p \delta (y^{\mu}-1) + (1-p) \delta (y^{\mu}+1), \label{ydistrib}
\end{equation}
where $p$ corresponds to the bias parameter. 
The encoder is simply defined as follows, 
\begin{equation}
  \mathcal{F}(\vec{y}) \equiv \mathop{\rm argmin}_{\hat{\vec{s}} \in \{-1,1\}^N} \,
  d(\vec{y},\mathcal{G}(\hat{\vec{s}})). \label{encoder}
\end{equation}
\par
Next, to decode the compressed message $\vec{s}$ 
we make use of the already introduced tree-like perceptrons. 
As in the error correcting code scheme, we prepare a set of $M$ input vectors 
$(\vec{x}^1,\ldots,\vec{x}^{\mu},\ldots,\vec{x}^M)$ 
which are drawn independently and uniformly on $\{ -1,1 \}$. This will play the role of the codebook. 
The compressed message $\vec{s}$ is used as the coupling vector of the network. 
Then, each input vector $\vec{x}^{\mu}$ is fed sequentially 
into the network generating a corresponding scalar $\hat{y}^{\mu}$ 
at the output of the network finally resulting in a $M$-dimensional vector $\hat{\vec{y}}$. 
This gives us the reconstructed message 
which should satisfies $E[\frac 1 M d(\vec{y},\hat{\vec{y}})] \leq D$ 
where $D$ is the desired fidelity criterion which measure the amount of error 
between the reconstructed message $\hat{\vec{y}}$ and the original message $\vec{y}$. 
\par
To keep notation as general as possible, 
as long as explicit use of the decoder is not necessary in computations, 
we will again denote the transformation perform on the vector $\vec{s}$ 
by the respective tree-like perceptrons using the notation
$\mathcal{F}_k ( \sqrt{\frac K N} \vec{s}_l \cdot \vec{x}_l^{\mu} )$. 
\par
The encoding phase can be viewed as a classical perceptron learning problem, where one tries 
to find the weight vector $\vec{s}$ which minimizes the distortion function $d(\vec{y},\hat{\vec{y}})$ for 
the original message $\vec{y}$ and the random input vector $\vec{x}$. 
The vector $\vec{s}$ which achieve this minimum gives us the codeword to be send to the decoder. 
Therefore, in the case of a lossless compression scheme(i.e.: $D=0$), 
evaluating the rate distortion property of the present scheme is equivalent 
to finding the number of couplings $\vec{s}$ which satisfies 
the input/output relation $\vec{x}^{\mu} \mapsto y^{\mu}$. In other words, this is equivalent 
to the calculation of the storage capacity of the network \cite{Gardner1988,Krauth1989}. 
\par
The typical performance of this scheme was already studied using the Replica Method (RM) \cite{Cousseau2008} 
and it was shown that each of the three proposed network %[MODIFIED]
can reach the optimal Shannon bound 
at the infinite codeword length limit (when $N \to \infty$ and $M \to \infty$ 
while the code rate $R$ is kept finite) under some specific condition. 
\par
The PTH and the CTH were shown to reach the Shannon bound for any number of hidden units $K$ 
(any odd number of hidden units in the case of the CTH) 
if the threshold parameter $k$ of the non-monotonic transfer function is properly tuned. 
The CTO was shown to reach the Shannon bound 
when its number of hidden units $K$ becomes infinite and with a properly tuned threshold parameter $k$ only. 

%%%%%%%%%%%%%%%%%%%%%%%%%%%%%%%%%%%%%%%%%%%%%%%%%%%%%%%%%%%%%%%%%%%%%%%%%%%%%%%%%%%%%%%%%%%%%
\section{Belief-propagation-based algorithms}
\label{sec:BP}

\par
In this section we briefly introduce the BP algorithm 
and how it can be used to infer an approximation of the marginalized posterior probabilities. 
The BP or sum-product algorithm is originally designed to compute exact marginalization 
on a factor graph which is a tree. However it is known to give very good performance 
even for non tree factor graph in various cases. 
For a formal introduction of the BP algorithm, see \cite{Opper2000, MacKay2003}.
\par
So far, the BP algorithm was already applied by Hosaka et al. 
in the case of lossy compression using the simple perceptron \cite{Hosaka2006}, 
but not in the case of the MLP. %[MODIFIED]
We follow the footsteps of their work to investigate 
both of the BP decoder for error correcting code and the BP encoder for lossy compression, 
which are based on multilayer perceptrons. %[MODIFIED]
It should be noted that we can discuss a basic BP algorithm 
for error correcting codes and lossy compression at a time. %[ADDED]
For the BP to be used, we need to have a factorizable probability distribution. 
Based on the statistical mechanical framework used in the references \cite{Cousseau2008bis, Cousseau2008}, 
the posterior probability of each case 
(either in the error correcting code scheme and lossy compression scheme) 
can be represented by a Boltzmann distribution 
\begin{equation}
  p(\vec{s}|\vec{y},\{\vec{x}\};\beta) 
  = \frac {\exp [-\beta \mathcal{H}(\vec{s},\vec{y},\{\vec{x}\}) ]}
          {Z(\vec{y},\{\vec{x}\};\beta)} ,
\end{equation}
where $\mathcal{H}(\vec{s},\vec{y},\{\vec{x}\})$ denotes the relevant Hamiltonian 
and $Z(\vec{y},\{\vec{x}\};\beta)$ the relevant partition function. 
The notation $\{ \vec{x} \}$ denotes the fact that the random vectors $\vec{x}^{\mu}$ 
are already fixed and known, which are random quenched variables. 
\par
\begin{figure}[t]
  \vspace{0mm}%<--space
  \begin{center}
  \includegraphics[width=0.7\linewidth,keepaspectratio]{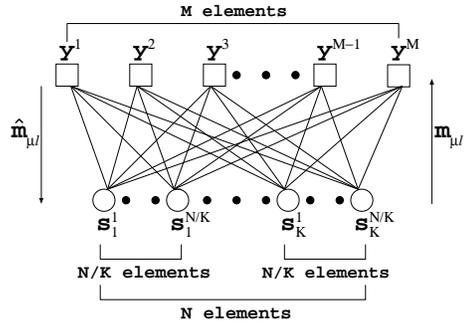}
  \end{center}
  \caption{Factor graph of the Boltzmann distribution}
  \label{fig:factorGraph}
  \vspace{0mm}%<--space
\end{figure}

In order to use the BP algorithm, this Boltzmann distribution %[DELETED] (we assume that)
can be factorized such that the Boltzmann factor can be decomposed into 
\begin{equation}
  \exp [ -\beta \mathcal{H}(\vec{s},\vec{y},\{\vec{x}\}) ] = 
  \prod_{\mu=1}^M G_{k,\mu} \biggl( \biggl\{ \sqrt{\frac {K} {N}} 
  \vec{s}_l \cdot \vec{x}^{\mu}_l \biggr\} \biggr) ,
\end{equation}
where the expression of the function $G_{k,\mu}$ depends on the scheme considered. 
In Appendix \ref{appendix.A}, the derivaton of the BP-based decoders for error correcting codes are given. %[ADDED]
In Appendix \ref{appendix.B}, the BP-based encoders for lossy compression are derived. %[ADDED]
Following from this assumption, we can write down the factor graph representation 
of the Boltzmann distribution as a bipartite graph (Figure \ref{fig:factorGraph}), 
\par
In the BP, it is assumed that the secondary contribution 
of a single variable $s^i_l$ or $y^\mu$ is small and must be neglected. %[ADDED]
Under this assumption, the factor graph shown in Fig. \ref{fig:factorGraph} is regarded as having a tree-like architecture. %[ADDED]
Now let us write down the set of messages flowing 
from the source sequence to the codeword and vice versa. 
We then have the following equations: 
\begin{eqnarray}
  & & 
  \hat{\rho}^t_{\mu i l} (s^i_l) 
  = \sum_{\vec{s} \backslash \{ s^i_l \} } 
  G_{k,\mu} \biggl( \left\{ \sqrt{\frac {K} {N}} 
  \vec{s}_l \cdot \vec{x}^{\mu}_l \right\} \biggr) \nonumber \\
  & &
  \qquad \qquad 
  \times \biggl( \prod_{i^{\prime} \neq i}^{N/K} 
  \rho^t_{\mu i^{\prime} l} (s^{i^{\prime}}_l) \biggr) 
  \biggl( \prod_{l^{\prime} \neq l}^{K} \prod_{i^{\prime} =1}^{N/K} 
  \rho^t_{\mu i^{\prime} l^{\prime}} (s^{i^{\prime}}_{l^{\prime}}) \biggr) , \\
  & & 
  \rho^{t+1}_{\mu i l} (s^i_l) 
  = C_{\mu i l}  q^t_{i l} (s^i_l)
  \biggl( \prod_{\mu^{\prime} \neq \mu}^{M} \hat{\rho}^t_{\mu^{\prime} i l} (s^i_l) \biggr),
\end{eqnarray}
where $C_{\mu i l}$ denotes the relevant normalization constant 
and $q^t_{i l} (s^i_l)$ denotes the prior. 
$\hat{\rho}^t_{\mu i l} (s^i_l)$ denotes the message received by the random variable $s^i_l$ 
from the source sequence bit $y^\mu$ at time step $t$. 
$\rho^{t+1}_{\mu i l} (s^i_l)$ denotes the message sent 
by the random variable $s^i_l$ to the source sequence bit $y^\mu$ at time step $t+1$.
At time $t+1$, the pseudo posterior marginals is given as 
\begin{eqnarray}
  & & p^{t+1} (s^i_l | \vec{y} , \{ \vec{x} \} ; \beta) 
  = \sum_{\vec{s} \backslash \{ s^i_l \} } 
  p^{t+1} (\vec{s} | \vec{y} , \{ \vec{x} \} ; \beta) \nonumber \\
  & & 
  \qquad \qquad \qquad \quad 
  \approx C_{i l}  q^t_{i l} (s^i_l)
  \left( \prod_{\mu = 1}^{M} \hat{\rho}^t_{\mu i l} (s^i_l) \right) ,
\end{eqnarray}
where $C_{i l}$ denotes the relevant normalization constant. 
\par
We obtain the BP-based algorithm as follows: 
\begin{eqnarray}
  m^{t+1}_{il} 
  = \tanh \bigg[ \sum_{\mu=1}^M \sqrt{\frac K N} x^{\mu}_{il} \Phi^t_{k,\mu l}
  + m^t_{il} \mathfrak{G}^t_{k,l} + \frac 1 2 \ln \frac {q^t_{i l} (1)} {q^t_{i l} (-1)} \bigg] , \label{refineBP-text}
\end{eqnarray}
where we have inserted back the term depending on the prior 
and we put $p^t (s^i_l | \vec{y} , \{ \vec{x} \} ; \beta) = \frac 12 (1+ m^t_{i l} s^i_l)$. 
Detail of calculation and definitions both $\Phi^t_{k,\mu l}$ and $\mathfrak{G}^t_{k,l}$ are available in Appendix \ref{appendix.A}. 
The MPM estimator at time step $t$ is given by 
\begin{eqnarray}
s^i_l = \sgn (m^{t}_{il}) . \label{refineMPM-text}
\end{eqnarray}
This BP algorithm requires $O(N^2)$ operations for each step.

%%%%%%%%%%%%%%%%%%%%%%%%%%%%%%%%%%%%%%%%%%%%%%%%%%%%%%%%%%%%%%%%%%%%%%%%%%%%%%%%%%%%%%%%%%%%%
\section{Empirical performance}
\label{sec:performance}

%%%%%%%%%%%%%%%%%%%%%%%%%%%%%%%%%%%%%%%%%%%%%%%%%%%%%%%%%%%%%%%%%%%%%%%%%%%%%%%%%%%%%%%%%%%%%
\subsection{Error correcting code case}

\par
In this section we show the results we obtain by using the BP algorithm 
as a decoder of the scheme. 
\par
In the case of error correcting codes, 
the Edwards-Anderson order parameter $q$ is $q=1$ in the ferromagnetic phase, 
implying that $| \left\langle \vec{s}_l \right\rangle |^2=1$, 
where $\left\langle \ldots \right\rangle$ denotes the average with respect to $\vec{y}$ and $\vec{x}$. 
This means that a simple uniform prior can be used efficiently 
and there is no uncertainty about the sign of $s^i_l$. 
However, as it will be discussed further with the lossy compression case, 
we introduce a more refine prior, so-called an inertia term, of the following form 
\begin{eqnarray}
  q^t_{il} (s^i_l) & = & e^{s^i_l \tanh^{-1} (\gamma m^t_{il})},
\end{eqnarray}
where $0 \leq \gamma < 1$ denotes an amplitude of the inertia term. %[MODIFIED]
Note that $\gamma$ is set by trial and error. %[ADDED]
This method was already successfully applied by Murayama \cite{Murayama2004} 
for a lossy compression scheme. 
In the sequel, if nothing is explicitly precised about $\gamma$, 
then it means that we used $\gamma=0$, corresponding to the simple uniform prior.
\par
The general procedure is as follows. 
In each case, the threshold parameter $k$ is set to the optimal theoretical value. 
First, an original message $\vec{s}^0$ is generated from the uniform distribution. 
Then the original message is turned into a codeword $\vec{y}_0$ using the relevant network. 
The codeword is then fed into the binary asymmetric channel 
where it is corrupted by noise according to the parameters $p$ and $r$. 
The decoder receives the corrupted codeword $\vec{y}$ at the output of the channel. 
The BP is finally used to infer back the original message $\vec{s}^0$ 
using the corrupted codeword $\vec{y}$. 
The BP-based decoders are shown in Appendix \ref{appendix.A}. 
\par
We conducted two types of simulations. 
In the first one, the number of hidden units $K$, 
the size of the original message $N$, 
and the parameters $(p,r)$ of the BAC are kept constant. 
The changing parameter is the size of the codeword $M$ 
which results in different values for the code rate $R=N/M$. 
For each value of $R$ tested, we perform $100$ runs. 
For each run, we perform $100$ BP iterations 
and the resulted estimated message $\vec{s}$ is compared with the original one $\vec{s}^0$ 
using the overlap value $\frac 1 N \vec{s} \cdot \vec{s}^0$. 
The code rate is plotted against the mean value of the overlap. 
The author are well aware that in general, 
information theorists plot the performance of an error correcting code scheme 
using error probability plot in logarithmic scale. 
However, the present BP calculations still requires 
a computational cost of order $O(N^2)$ which prevent such drawing to be feasible. 
On top of that, the author believes that the main interest 
of the present schemes at the present state of research is 
from a theoretical point of view rather than a practical point of view. 
The performance plot intends to give an general idea 
about the typical performance obtained using the BP with these schemes 
but does not aim at discussing possible practical implementation of these schemes. 
We believe the performance exhibited by these schemes 
at the present time to be too limited to be worth such discussion. 
\par
In the second type of experiment, 
we try to shed light on the structure of the solution space. 
For this purpose, we fix the value of $K,N,M,p,r$ and generate 
an original message $\vec{s}^0$. 
We let run the BP algorithm 
and get a estimated message $\vec{s}$ after $100$ iterations. 
Then we keep the same original message 
and let run the BP again but with different initial values. 
After $100$ iterations we get another estimated message $\vec{s}^{\prime}$. 
We perform the same procedure $30$ times 
and we calculate the average overlap $\frac 1 N \vec{s} \cdot \vec{s}^{\prime}$ 
between all the obtained estimated messages. 
Next we generate a new original message $\vec{s}^0$ and 
do the same procedure for $50$ different original messages. 
We finally plot the obtained overlap using histograms, 
thus reflecting the distribution of the solution space.

%%%%%%%%%%%%%%%%%%%%%%%%%%%%%%%%%%%%%%%%%%%%%%%%%%%%%%%%%%%%%%%%%%%%%%%%%%%%%%%%%%%%%%%%%%%%%
\subsubsection{Parity tree with non-monotonic hidden units (PTH)}

\par
\begin{figure}[t]
  \vspace{0mm}%<--space
  \begin{center}
  \includegraphics[width=0.6\linewidth,keepaspectratio]{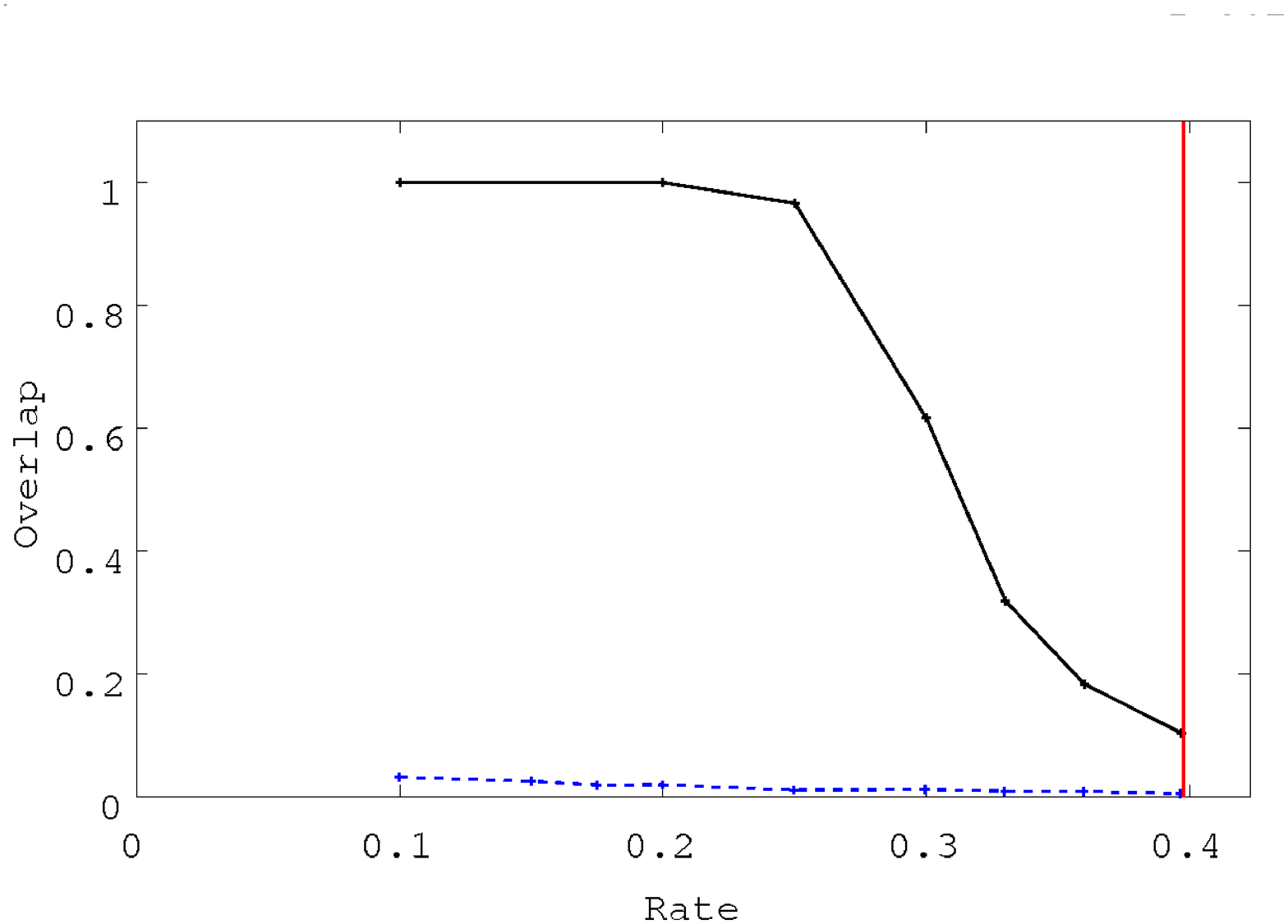}
  \end{center}
  \caption{Empirical performance of the BP-based decoder for error correcting codes using the PTH with $K=1$ (solid) and $K=3$ (dashed). 
    We set $p=0.1$, $r=0.2$ and $\gamma=0$ (set by trial and error) and used $N=1000$ (for $K=1$) and $N=999$ (for $K=3$). The vertical line represents the Shannon bound.}
  \label{fig:PTH_K1_3_Rate}
  \vspace{0mm}%<--space
\end{figure}

We show the results obtained for the PTH with $K=1$ and $K=3$ hidden units 
in Figure \ref{fig:PTH_K1_3_Rate}. 
The vertical line represents the Shannon bound, 
that is the theoretical limit for which decoding is still successful (i.e.: overlap is $1$). 
The average overlap for $100$ trials is plotted. 
While the Shannon bound gives a theoretical optimal code rate of $R \approx 0.4$, 
in this case for $K=1$, the performance of the BP starts to deteriorates rapidly for $R>0.25$. 
This shows limitation of the BP performance. 
We tested several configuration with different value for $p$ and $r$ (BSC case and Z channel case), 
and the general tendency is always the same. Far from the Shannon bound, the performance deteriorates rapidly.
\par
Next the same experiment with $K=3$ hidden units shows 
that the BP fails completely to decode the corrupted codeword. 
The average overlap is $0$ even for low value of $R$. 
We always got the same results for any value of $p$ and $r$. 
In fact, for any $K>1$, it seems that the BP always fails to converge to any relevant solution. 
This result is surprising and might indicates 
that the number of suboptimal states is so important that this prevent the BP to work.
\par
\begin{figure}[t]
  \vspace{0mm}%<--space
  \begin{center}
  \small
  \includegraphics[width=0.6\linewidth,keepaspectratio]{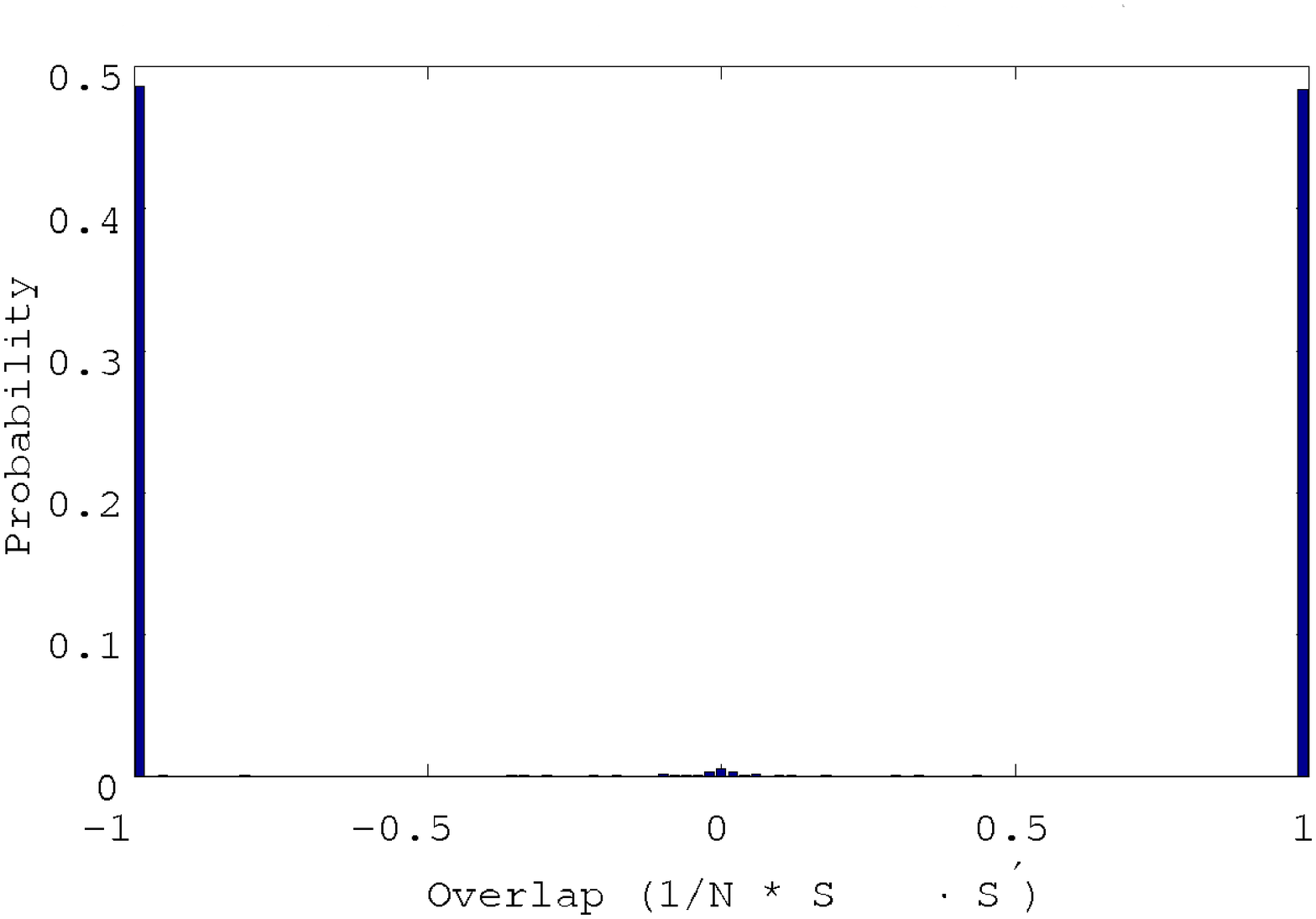}\\
  {\footnotesize (a) $K=1$, $N=1000$} \\
  \includegraphics[width=0.6\linewidth,keepaspectratio]{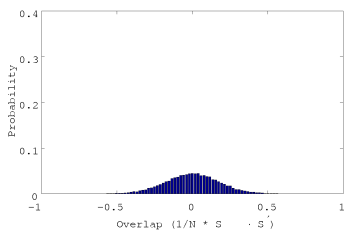}\\
  {\footnotesize (b) $K=3$, $N=102$} \\
  \includegraphics[width=0.6\linewidth,keepaspectratio]{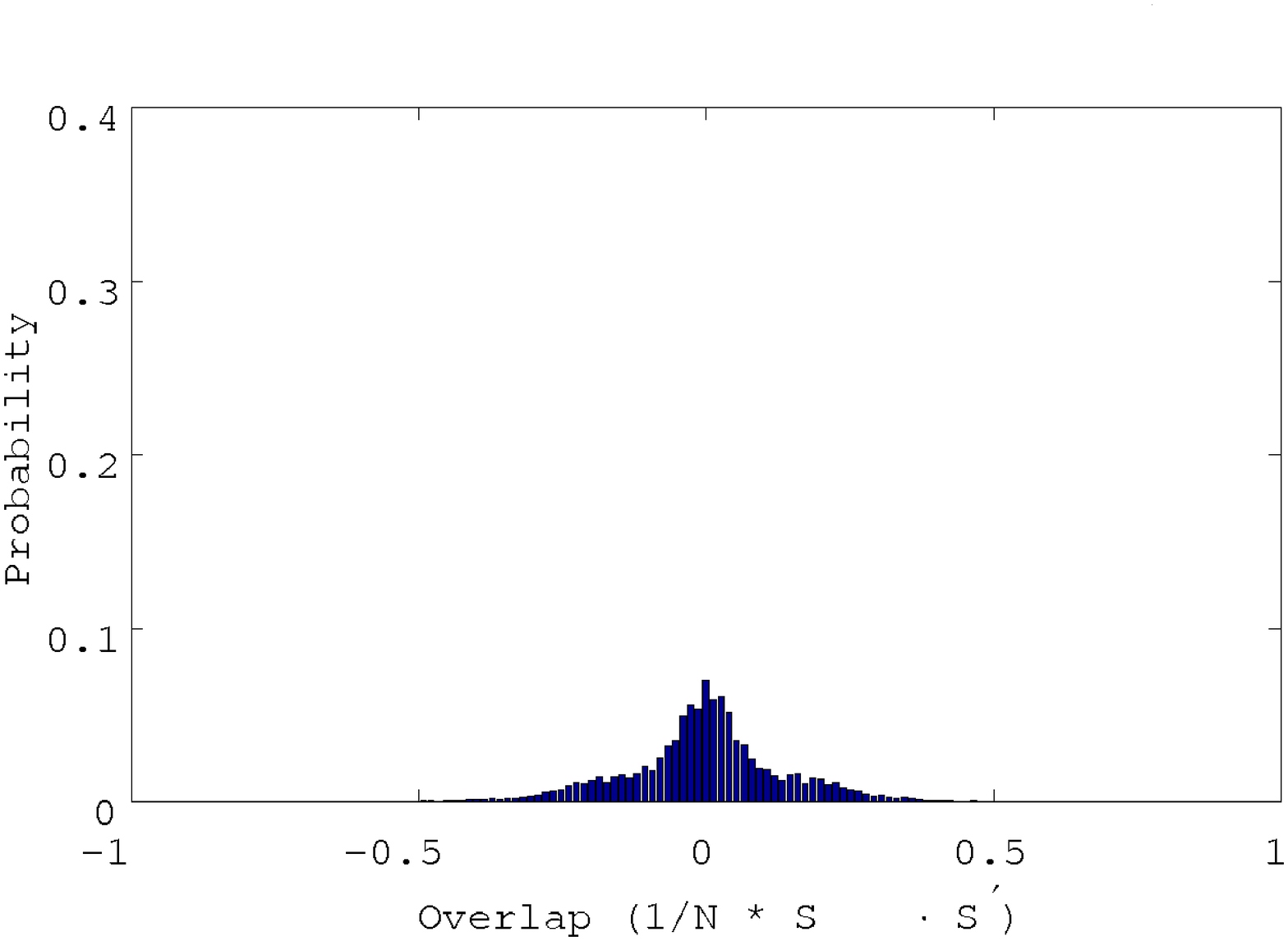} \\
  {\footnotesize (c) $K=3$, $N=1002$} 
  \end{center}
  \caption{Overlap of the solutions given by the BP-based decoder for error correcting codes using the PTH 
    with $R=0.25$, $p=0.1$, $r=0.2$ and $\gamma=0$. 
    (a) $K=1$ and $N=1000$. The empirical overlap with the original message is $0.97$.
    (b) $K=3$ and $N=102$. The empirical overlap with the original message is $0.08$.
    (b) $K=3$ and $N=1002$. The empirical overlap with the original message is $0.03$.}
  \label{fig:PTH_K1_N1000_Distrib}
  \vspace{0mm}%<--space
\end{figure}

Then we try to investigate the structure of the solution space. 
We plot the histograms of the overlap of the solutions obtained 
using the BP (when $K=1$) in Figure \ref{fig:PTH_K1_N1000_Distrib} (a).
In this case, we see that the BP converges to two different solutions 
with opposite sign which corresponds to $\pm \vec{s}^0$. 
This is normal and comes from the mirror symmetry of the function $f_k$. 
In this case the solution space is simple, 
with two dominant attractor given by $\vec{s}^0$ and $- \vec{s}^0$.
\par
Then we perform the same experiment but with $K=3$ and $N=102$. 
Results are plotted in Figure \ref{fig:PTH_K1_N1000_Distrib} (b). 
We obtain a Gaussian like distribution centered on $0$. 
This means that the solution given 
by the BP are almost uncorrelated between each others. 
They do not correspond to any relevant solution 
and the empirical overlap is almost $0$. 
We then conduct the same experiment keeping the code rate unchanged 
but for an original message of $1000$ bits. 
Results are shown in Figure \ref{fig:PTH_K1_N1000_Distrib} (c).
The distribution becomes sharper, centered on $0$, 
meaning that the solutions given by the BP are completely uncorrelated. 
The number of suboptimal states becomes very large 
and the BP completely fails to converge to a relevant solution.
\par
To conclude the case of the PTH, we can say that for $K=1$, 
the BP converges but with performance far from being Shannon optimal. 
For $K>1$, the BP completely fails. 
This is probably due to a rise of suboptimal states when using more than $1$ hidden unit.

%%%%%%%%%%%%%%%%%%%%%%%%%%%%%%%%%%%%%%%%%%%%%%%%%%%%%%%%%%%%%%%%%%%%%%%%%%%%%%%%%%%%%%%%%%%%%
\subsubsection{Committee tree with non-monotonic hidden units (CTH)}

\par
\begin{figure}[t]
  \vspace{0mm}%<--space
  \begin{center}
  \small
  \includegraphics[width=0.6\linewidth,keepaspectratio]{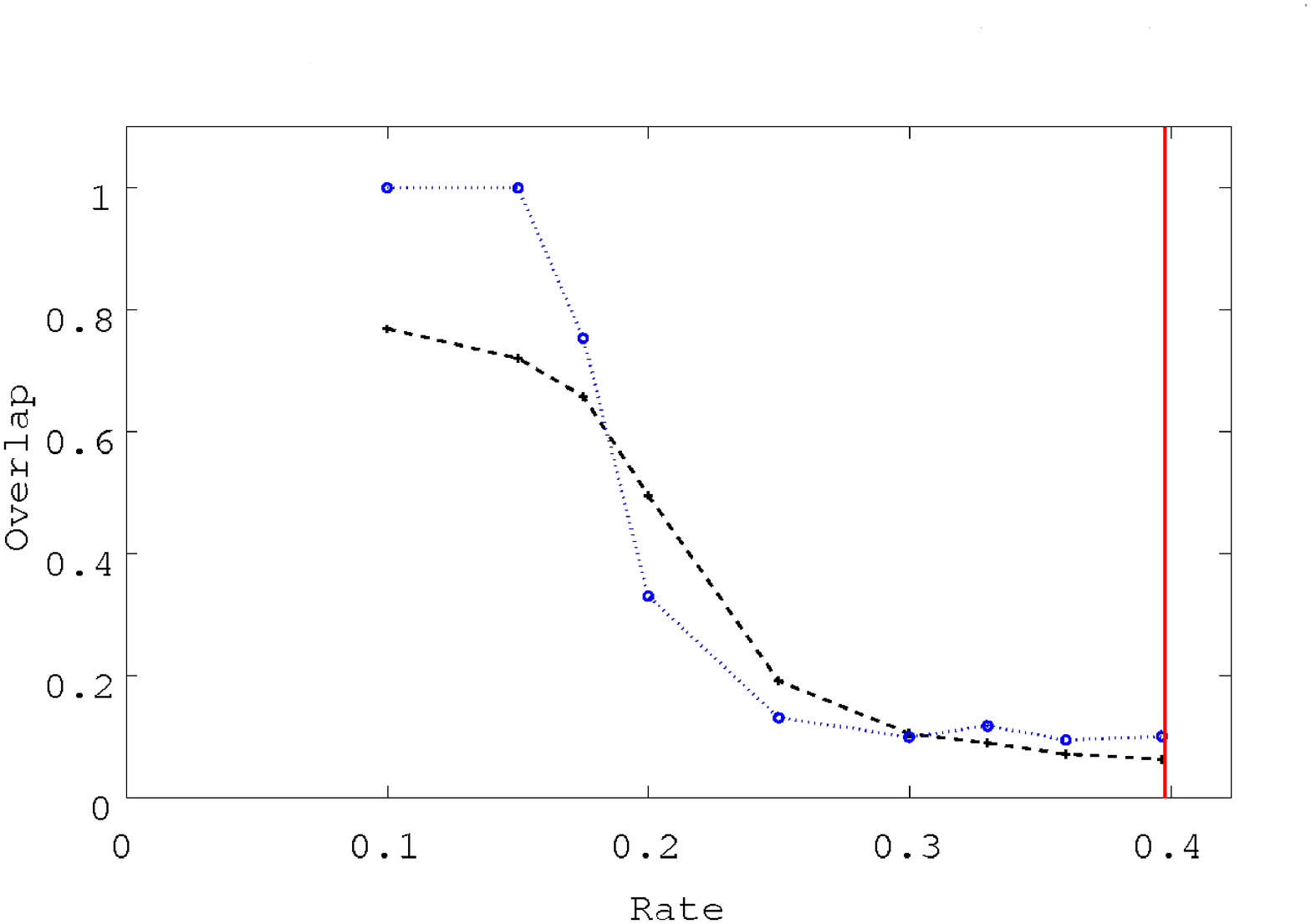} \\
  {\footnotesize (a) $K=3$, $N=999$, $\gamma\in\{0, 0.45\}$} \\
  \includegraphics[width=0.6\linewidth,keepaspectratio]{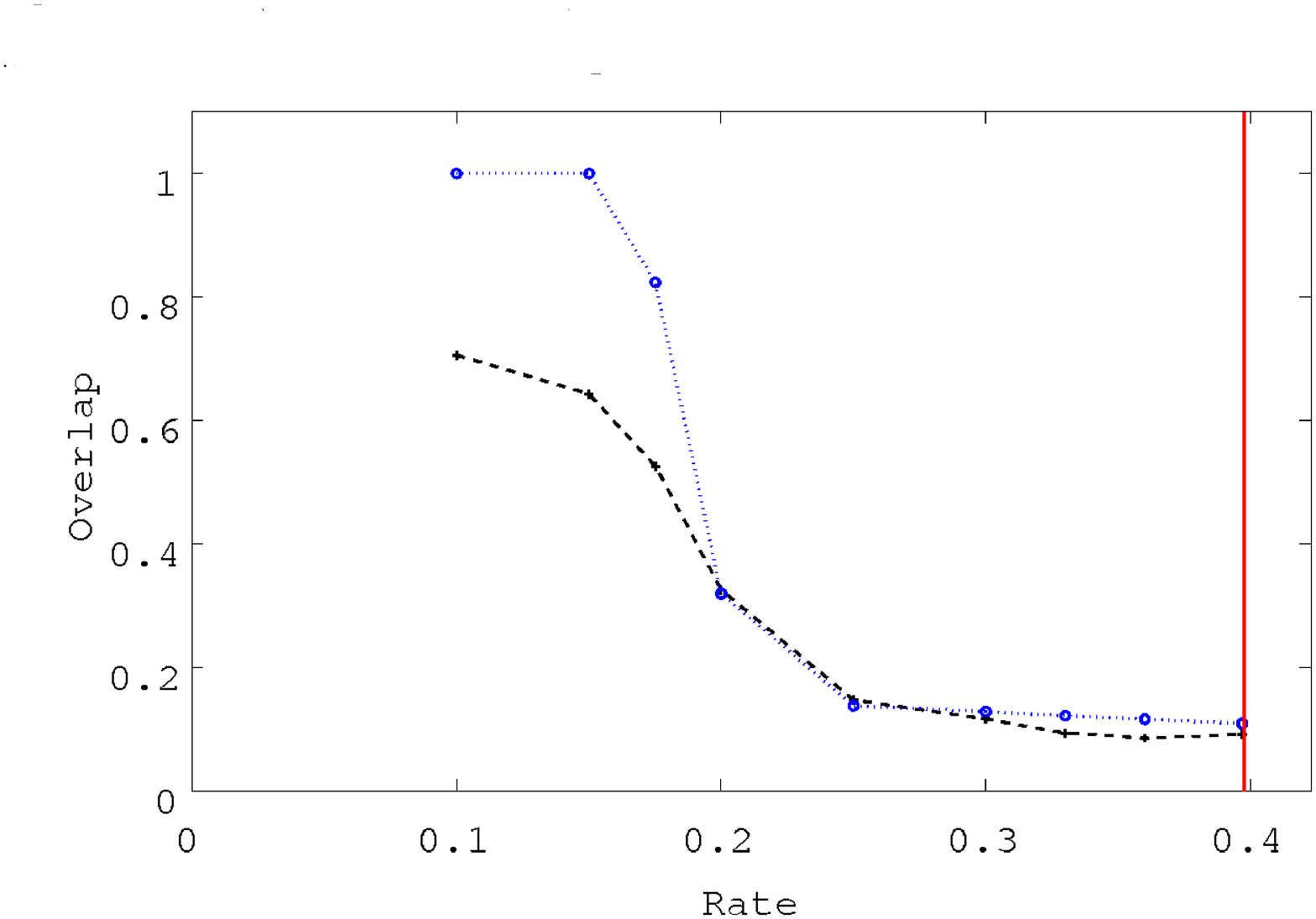} \\
  {\footnotesize (b) $K=5$, $N=1000$, $\gamma\in\{0, 0.45\}$} \\
  \includegraphics[width=0.6\linewidth,keepaspectratio]{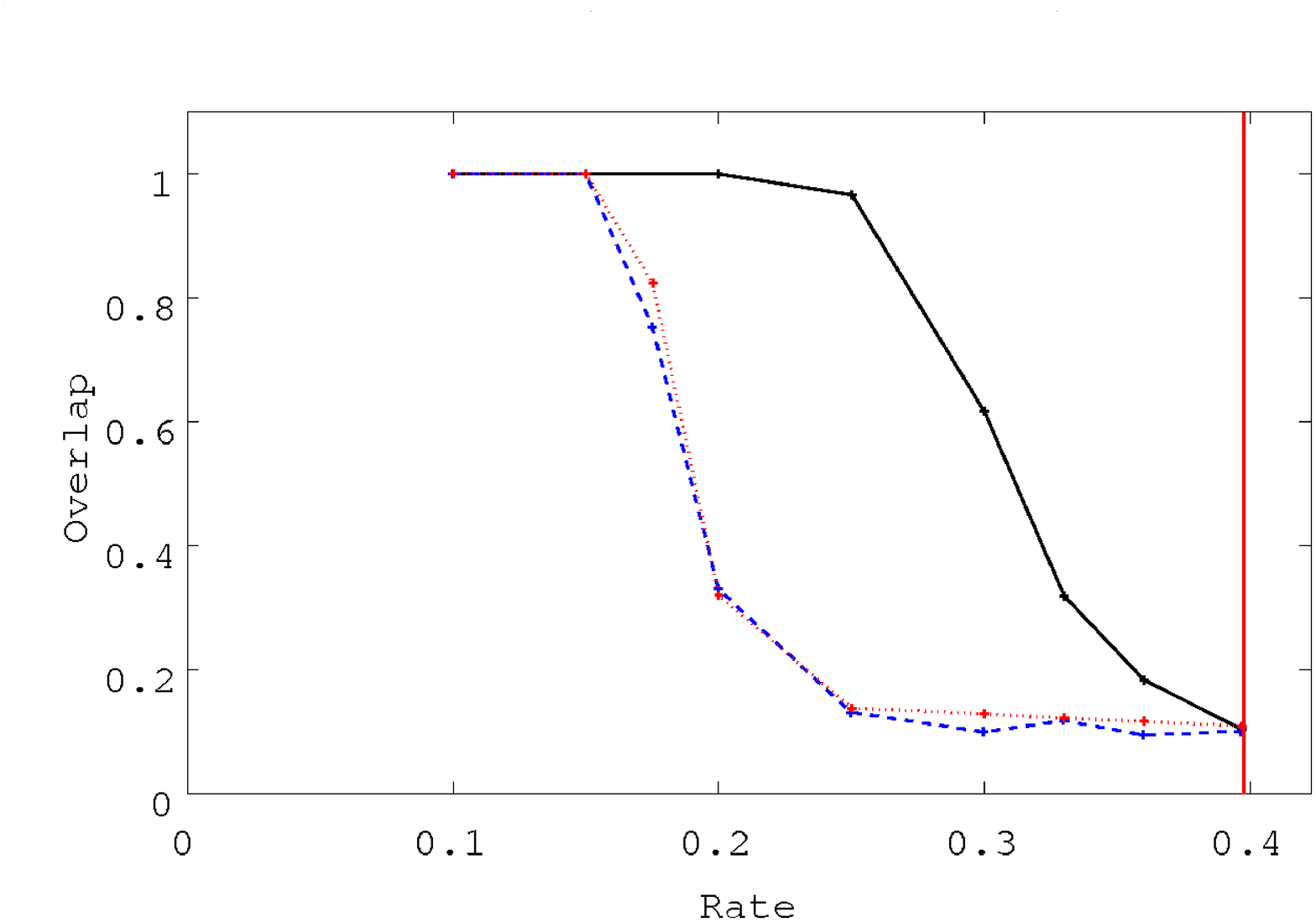} \\
  {\footnotesize (c) $K\in\{1,3,5\}$}
  \end{center}
  \caption{Empirical performance of the BP-based decoder for error correcting codes 
    using the CTH with $p=0.1$ and $r=0.2$. 
    The vertical line represents the Shannon bound. 
    (a) $K=3$ and $N=999$. The dashed line is for $\gamma=0$, the dotted line is for $\gamma=0.45$. 
    (b) $K=5$ and $N=1000$. The dashed line is for $\gamma=0$, the dotted line is for $\gamma=0.45$. 
    (c) $K=1$ (solid), $K=3$ (dashed) and $K=5$ (dotted) hidden units. 
    We set $N=1000$ for $K=1,5$ and $N=999$ for $K=3$. 
    We chose $\gamma=0$ for $K=1$ and $\gamma=0.45$ for $K=3,5$, which are set by trial and error. 
  }
  \label{fig:CTH_K3_Rate}
  \vspace{0mm}%<--space
\end{figure}

We show the results obtained for the CTH with $K=3$ hidden units 
in Figure \ref{fig:CTH_K3_Rate} (a). 
We do not show the result for $K=1$ 
because in this case, the CTH is equivalent to the PTH. 
The vertical line represents the Shannon bound. 
The average overlap for $100$ trials is plotted. 
\par
In this case it is interesting to note that for $\gamma=0$ and $R<0.15$, 
the BP fails to properly recover the original message 
but still seems to converge to some meaningful state. 
The average overlap is around $0.75$ but never reaches $1$. %[MODIFIED] but however -> but
This is probably due to local suboptimal attractors in the solution space. 
Adding a perturbation by inserting a non zero inertia term 
seems to be a good way to escape those suboptimal states 
and the best performance are obtained for $\gamma$ around $0.45$. 
However, for $R \le 0.15$ whatever the value $\gamma$ takes the performance quickly deteriorates. %[MODIFIED]
So the performance are very far from being optimal 
and suggest that the bigger the code rate is, 
the larger the number of suboptimal states are. 
\par
We then conduct the exact same experiment but for $K=5$. 
The results are shown in Figure \ref{fig:CTH_K3_Rate} (b) 
and are almost identical to the results obtain for $K=3$. 
We then make a comparison of the best performance obtained using the CTH. 
Results are shows in Figure \ref{fig:CTH_K3_Rate} (c). 
\par
The best performance are obtain for $K=1$. 
Increasing the number of hidden units clearly yields poorer performance. 
This is an interesting phenomenon and the only explanation is 
that the number of hidden units have a critical influence and the solution space structure. 
While theoretically any number of hidden units 
should be able to yield optimal performance, 
the BP clearly gives bad results for $K>1$. 
In a similar way as the study we have introduced in the first part of this paper. 
It is very likely that the intrinsic structure of MLPs is at the origin of this ill behavior. %[MODIFIED]
The number of hidden units seems to play a critical role 
in the organization of the solution space, 
and probably give rise to some complex geometrical features.
\par
\begin{figure}[t]
  \vspace{0mm}%<--space
  \begin{center}
  \small
  \includegraphics[width=0.6\linewidth,keepaspectratio]{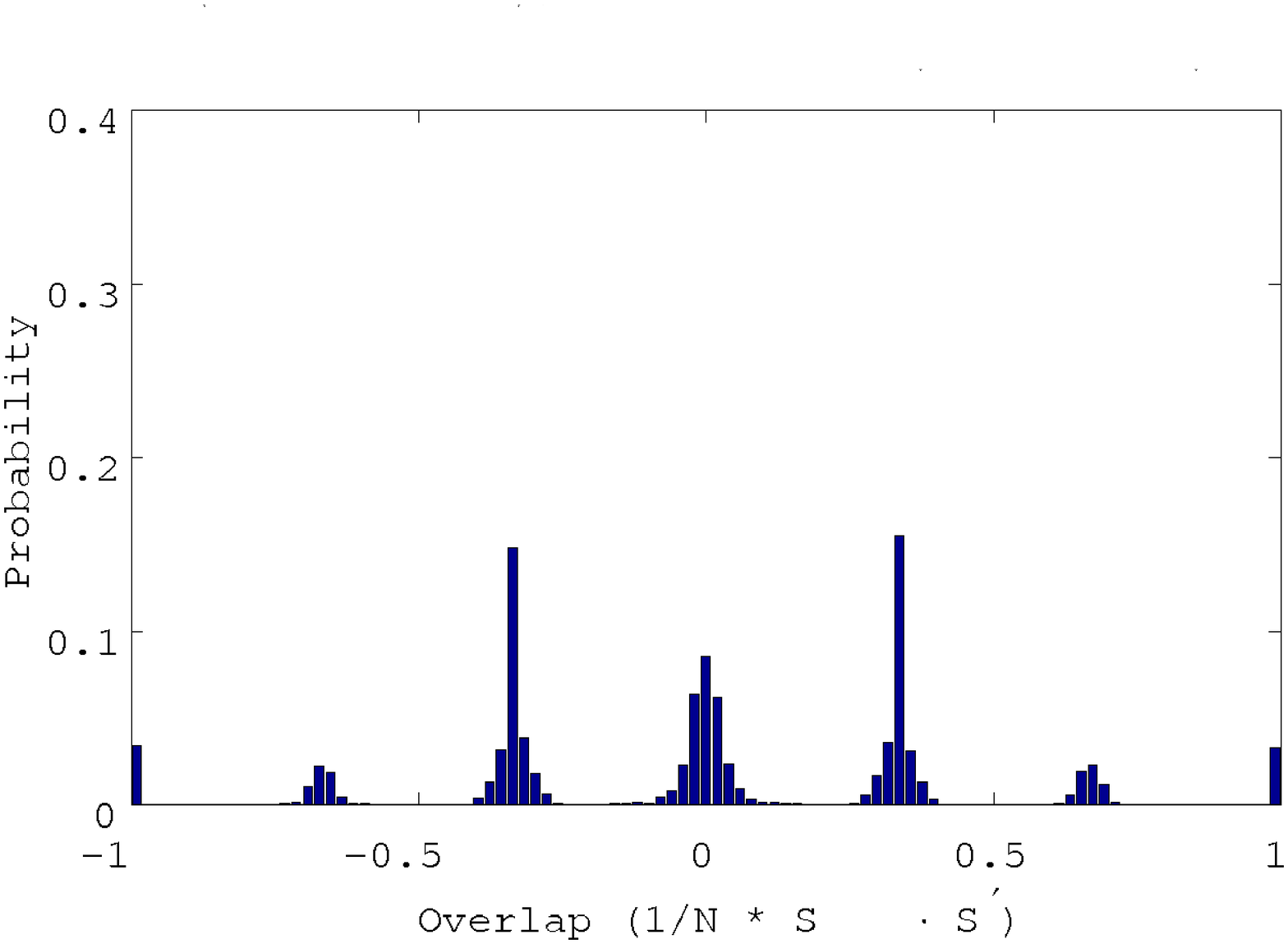} \\
  {\footnotesize (a) $\gamma=0$} \\
  \includegraphics[width=0.6\linewidth,keepaspectratio]{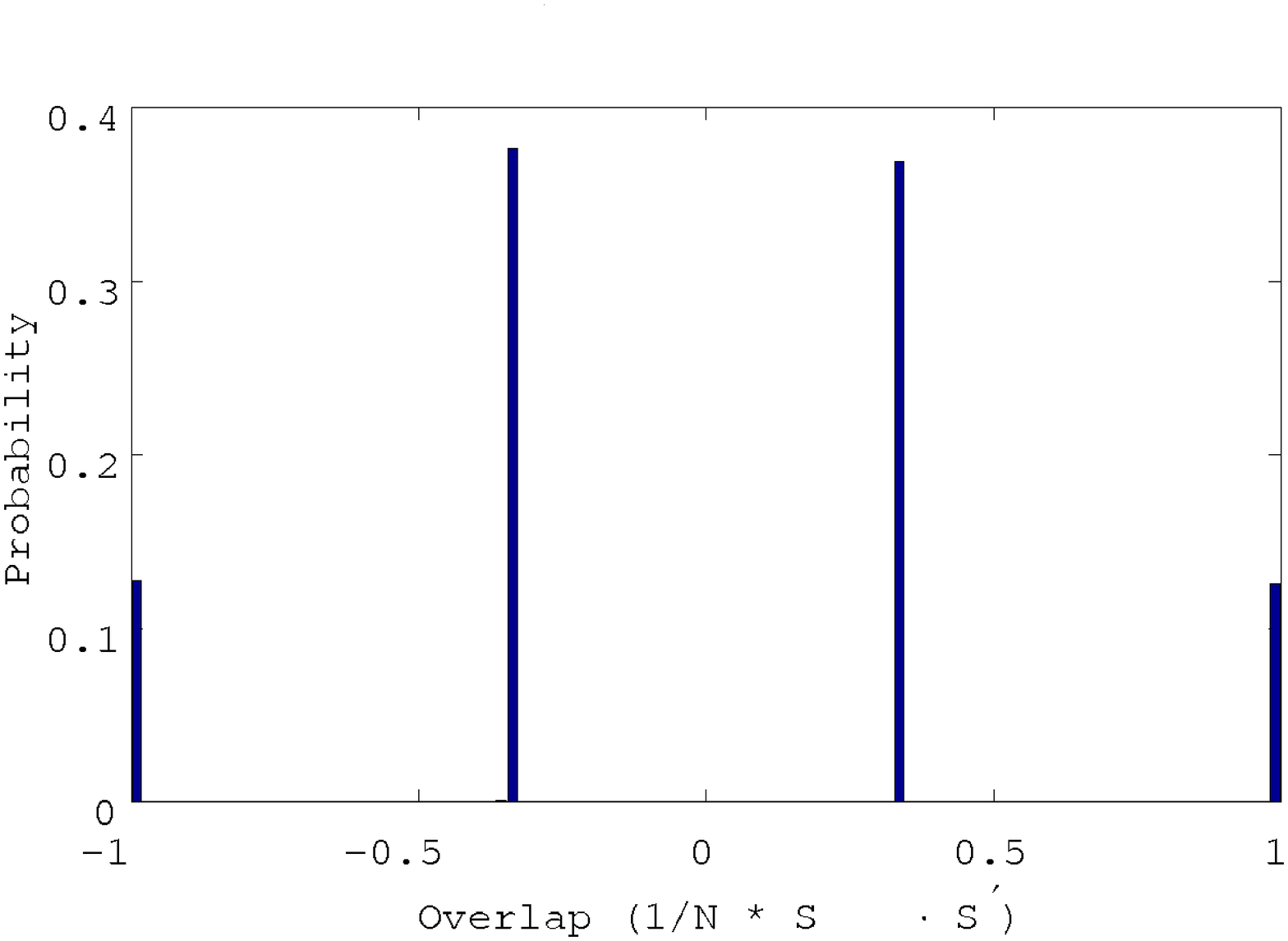} \\
  {\footnotesize (b) $\gamma=0.45$} 
  \end{center}
  \caption{Overlap of the solutions given by the BP-based decoder for error correcting codes using the CTH 
    with $K=3$, $N=999$, $M=6660$, $R=0.15$, $\gamma=0$, $p=0.1$ and $r=0.2$. 
    (a) $\gamma=0$. The empirical overlap with the original message is $0.74$. 
    (b) $\gamma=0.45$. The empirical overlap with the original message is $0.99$.}
  \label{fig:CTH_K3_N1000_R015_Distrib}
  \vspace{0mm}%<--space
\end{figure}

Then we try to investigate the structure of the solution space. 
First, we plot the histograms of the overlap of the solutions obtained 
using the BP with $K=3$, $N=999$ and $\gamma=0$ in Figure \ref{fig:CTH_K3_N1000_R015_Distrib} (a). 
Then we plot the histograms of the overlap of the solutions obtained 
using the BP with $K=3$, $N=999$ and $\gamma = 0.45$ 
in Figure \ref{fig:CTH_K3_N1000_R015_Distrib} (b). 
\par
For $\gamma=0$, we obtain seven peaks. 
Two tall peaks at $\pm 1/3$, one peak at $0$, 
and four small peaks at $\pm 2/3$ and $\pm 1$. 
The peaks located at $\pm 1$ and $\pm 1/3$ corresponds to successful decoding 
and reflect the possible combination of decoded messages when $K=3$. %[DELETED] "]" 
Indeed, because of the mirror symmetry in the CTH network, 
any combination of $\pm \vec{s}^0_l$ gives the same output. 
We therefore have a inherent indetermination on the original message, 
which can be easily removed by adding some simple header to the codeword.
\par
The two small peaks around $\pm 2/3$ corresponds to a partial success in decoding. 
Indeed, further investigation showed that those peaks correspond to codewords 
where two of the three $\vec{s}^0_l$ vectors have been successfully retrieved 
but the last vector was not. 
This means that the BP remained trapped in some local attractor, 
which probably depends on the initial values used by the BP. 
The interesting fact is that it affects only partially the BP performance 
in this case, showing that for the CTH the BP dynamics 
of each $\vec{s}_l$ is independent to the others to some extent. 
Finally, the peak around $0$ reflect a completely unsuccessful decoding. 
This explains the average overlap found of $0.74$. 
\par
%-----------------------------ADDED (FROM HERE)
In the $K=3$ system, for a given original message $\vec{s}^0$, 
eight messages 
$\{( \vec{s}^0_1,  \vec{s}^0_2,  \vec{s}^0_3),$ 
$( \vec{s}^0_1,  \vec{s}^0_2, -\vec{s}^0_3),$ 
$( \vec{s}^0_1, -\vec{s}^0_2,  \vec{s}^0_3),$ 
$( \vec{s}^0_1, -\vec{s}^0_2, -\vec{s}^0_3),$ 
$(-\vec{s}^0_1,  \vec{s}^0_2,  \vec{s}^0_3),$ 
$(-\vec{s}^0_1,  \vec{s}^0_2, -\vec{s}^0_3),$ 
$(-\vec{s}^0_1, -\vec{s}^0_2,  \vec{s}^0_3),$ 
$(-\vec{s}^0_1, -\vec{s}^0_2, -\vec{s}^0_3) \} \triangleq \mathcal{S}(\vec{s}^0)$, 
which includes the original message $\vec{s}^0=( \vec{s}^0_1,  \vec{s}^0_2,  \vec{s}^0_3)$, 
are mapped into a same codeword. 
So an additional $K$ bit information is necessary to specify the original message from the set $\mathcal{S}(\vec{s}^0)$. 
For instance, it is one of the additional information to add $1$ to each block $l$ 
as $\vec{s}^0_l \mapsto (1,\vec{s}^0_l)$ to specify the original message 
(the length of this information is negligible than the length of the original message). 
If the BP decoder correctly estimates the original message $\vec{s}^0$, 
the estimated message is identical to one of the element of the set $\mathcal{S}(\vec{s}^0)$ with equiprobability. 
Therefore, when the BP decoder estimates correctly, the histgram exhibits only four peaks located 
at $\pm 1$ (probability $1/8$) and $\pm 1/3$ (probability $3/8$). 
%-----------------------------ADDED (TO HERE)
\par
The case where $\gamma=0.45$ on the other hand, exhibits only four peaks 
located at $\pm 1$ and $\pm 1/3$. 
Therefore this means that decoding is always successful in this case as confirmed 
by the average overlap of $0.99$. 
This result shows that using a non-zero inertia term can be an efficient way 
of avoiding sub-optimal states by adding a small perturbation to the BP dynamics.
\par
To conclude for the CTH, it is very clear that using a number of hidden unit 
greater than $1$ is at the origin of some structural changes of the solution space, 
which provokes a dramatic performance drop. 
For $K=1$, successful decoding is ensured until $R=0.25$ while for $K=3$, 
successful decoding is ensured until $R=0.15$ only. 
However, between $K=3$ and $K=5$, we observe no substantial change. 
It seems that as $R$ increases, the suboptimal states' basin of attraction 
quickly becomes very large compared to the optimal solution one. 
The influence of the number of hidden units on the solution space geometry 
remains to be investigated in a future work.  

%%%%%%%%%%%%%%%%%%%%%%%%%%%%%%%%%%%%%%%%%%%%%%%%%%%%%%%%%%%%%%%%%%%%%%%%%%%%%%%%%%%%%%%%%%%%%
\subsubsection{Committee tree with a non-monotonic output unit (CTO)}

\par
\begin{figure}[t]
  \vspace{0mm}%<--space
  \begin{center}
  \includegraphics[width=0.6\linewidth,keepaspectratio]{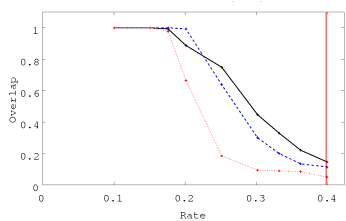}
  \end{center}
  \caption{Empirical performance of the BP-based decoder for error correcting codes using the CTO with $K=2$ (solid), $K=3$ (dashed) and $K=5$ (dotted). 
    We set $p=0.1$, $r=0.2$ and $\gamma=0$ (set by trial and error) 
    and used $N=1000$ (for $K=2$ and $K=5$), $N=999$ (for $K=3$). 
    The vertical line represents the Shannon bound.}
  \label{fig:CTO_K2_3_5_Rate}
  \vspace{0mm}%<--space
\end{figure}

We show the results obtained for the CTO with $K=2$, $K=3$ and $K=5$ hidden units 
in Figure \ref{fig:CTO_K2_3_5_Rate}. 
We do not show the result for $K=1$ because the CTO cannot be defined in this case. 
The vertical line represents the Shannon bound. 
The average overlap for $100$ trials is plotted.
For $K=2$ the BP successfully decodes the corrupted codeword until $R=0.17$ 
only and beyond this value the performance gradually decreases. 
Compared to the PTH/CTH with $K=1$, the CTO performance are poorer. 
However, this is not very surprising because as mentioned 
during the analytical study of the CTO case, 
the CTO is expected to reach the Shannon bound for an infinite number of hidden units only. 
The fact that we get suboptimal performance for finite $K$ is therefore not surprising. 
For $K=3$, the average performance is better and decoding is successful until $R=0.2$. 
In this case, it is worth using an extra unit. 
However, for $K=5$, the performance deteriorates 
and we get poorer performance than $K=1$. 
Nonetheless, the overall performance is still better than the CTH 
with the same number of hidden units.
\par
\begin{figure}[t]
  \vspace{0mm}%<--space
  \begin{center}
  \small
  \includegraphics[width=0.6\linewidth,keepaspectratio]{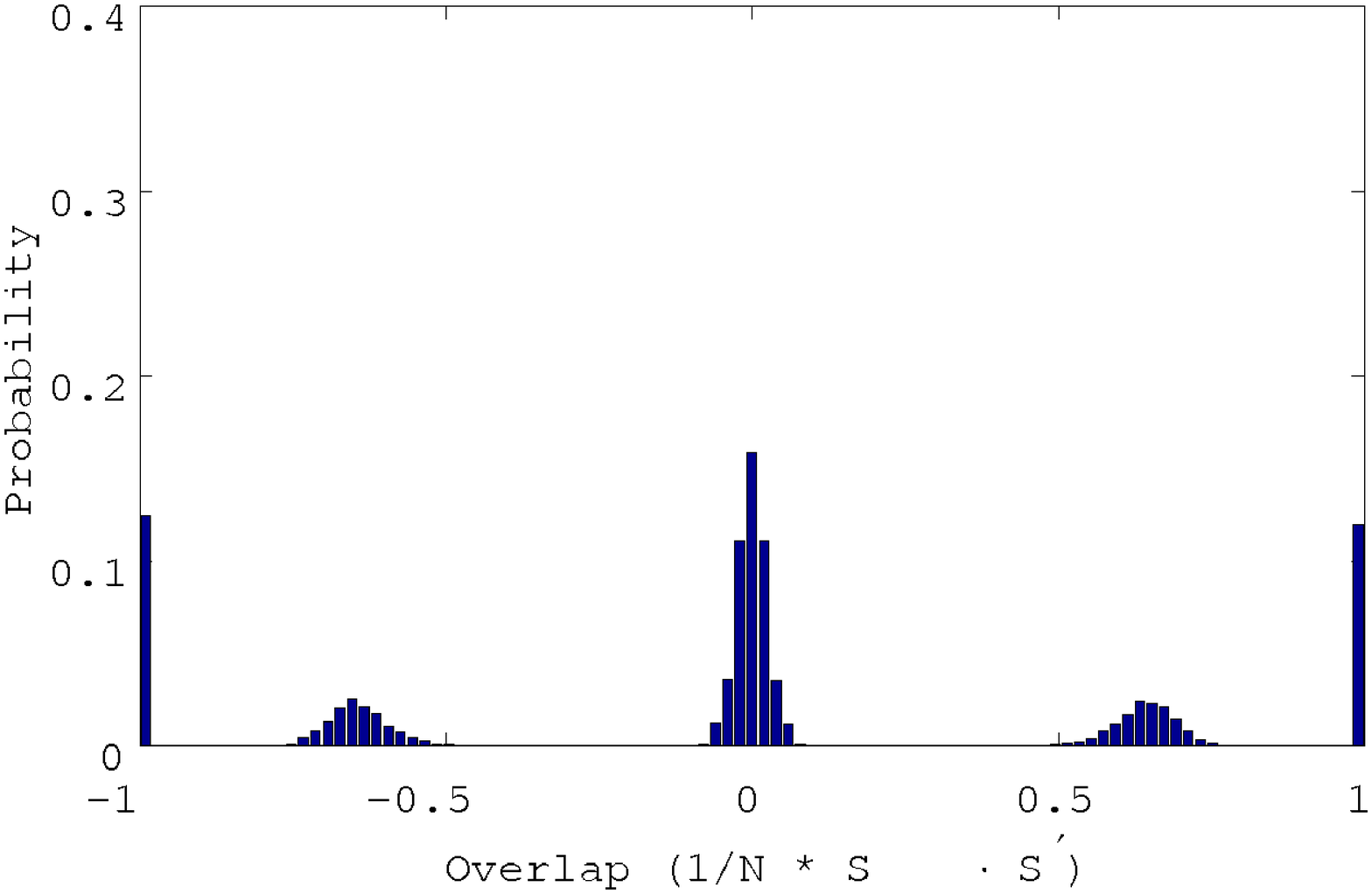} \\
  {\footnotesize (a) $K=2$, $N=1000$} \\
  \includegraphics[width=0.6\linewidth,keepaspectratio]{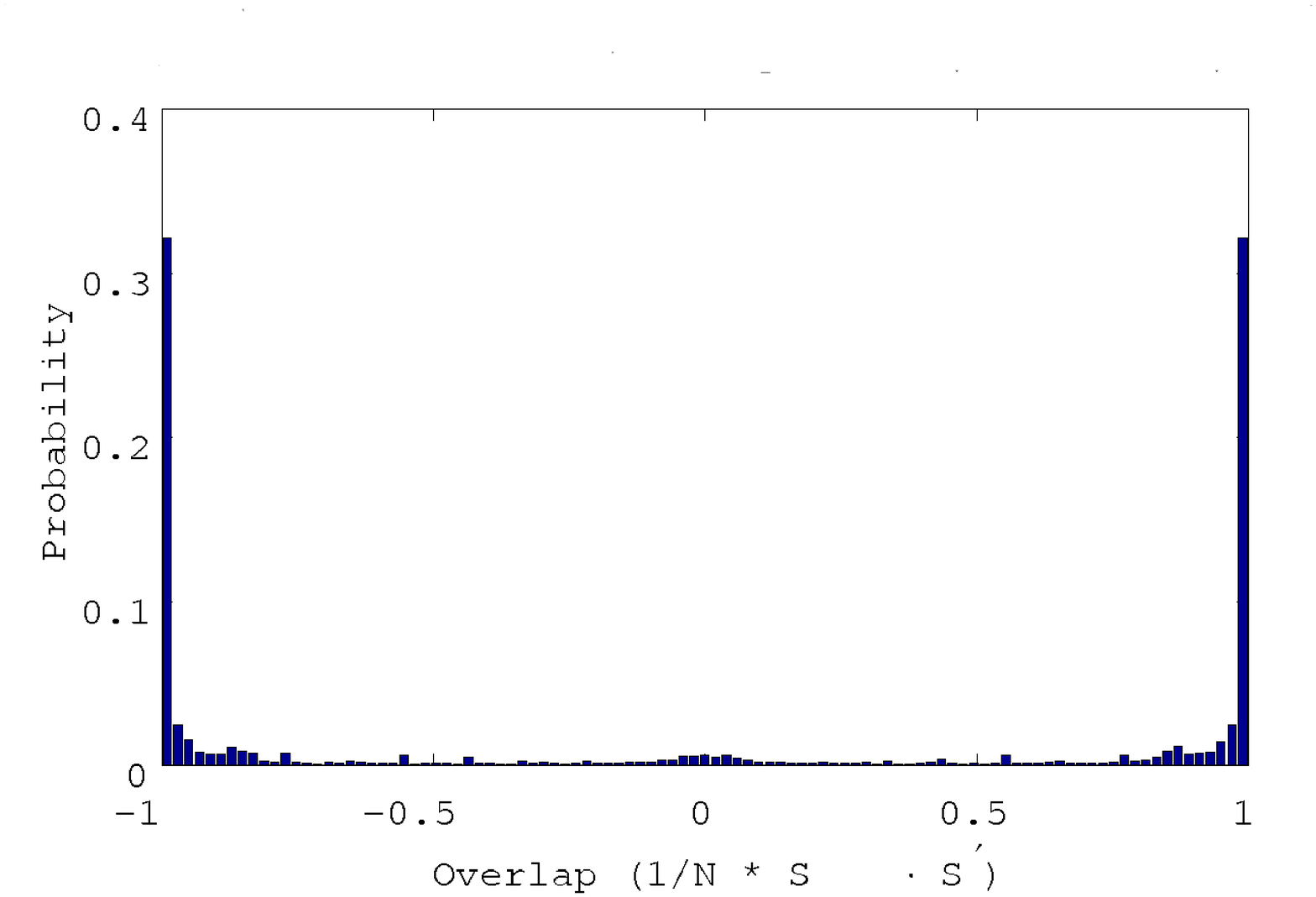} \\
  {\footnotesize (b) $K=3$, $N=999$}
  \end{center}
  \caption{Overlap of the solutions given by the BP-based decoder for error correcting codes using the CTO 
    with $K=2$, $N=1000$, $M=4000$, $R=0.25$, $\gamma=0$, $p=0.1$ and $r=0.2$. 
    (a) $K=2$ and $N=1000$. The empirical overlap with the original message is $0.76$. 
    (b) $K=3$ and $N=999$. The empirical overlap with the original message is $0.63$.}
  \label{fig:CTO_K2_N1000_Distrib}
  \vspace{0mm}%<--space
\end{figure}

Then we try to investigate the structure of the solution space. 
We plot the histograms of the overlap of the solutions obtained 
using the BP with $K=2$ and $N=1000$ in Figure \ref{fig:CTO_K2_N1000_Distrib} (a). 
We obtain three sharp peaks at $\pm 1$ and $0$ 
and two small peaks around $\pm 2/3$. 
The three sharp peaks correspond to successful decoding. 
As in the CTH case, their positions correspond to the possible combination 
of $\pm \vec{s}^0_l$. 
However, the other two small peaks corresponds 
to suboptimal states (the average overlap is $0.76$) 
and it is unclear what the value $\pm 2/3$ denotes. 
It might corresponds to some particular local attractor 
which should be investigated in the future. 
\par
Next we plot the histograms of the overlap of the solutions obtained 
using the BP with $K=3$ and $N=999$ in Figure \ref{fig:CTO_K2_N1000_Distrib} (b). 
For $K=3$, we obtain two sharp peaks at $\pm 1$ 
and a rather flat distribution connecting them (with a small concentration around $0$). 
The two sharp peaks correspond to successful decoding, 
while the rest of the distribution indicates suboptimal states. 
However, here there is no particular suboptimal states as in the case when $K=2$. 
This particularity is interesting and remains to be investigated.
\par
To conclude the case of the CTO, 
we can say that the BP reaches optimal performance for $K=3$ 
but decoding is still far from being Shannon optimal. 
However, as the analytical study already mentioned, 
the CTO is expected to yield Shannon performance 
when using an infinite number of hidden units so it is a little bit hard 
to explain the results of this section. 
Nevertheless, one may expect the performance to get better and better 
a K increases but this is not the case as denoted by the case when $K=5$. 
As for the other networks, it is very likely 
that the solution space exhibits strange geometrical features 
for $K>1$ (explaining the rise of suboptimal states), 
preventing the BP to converge properly for large values of $K$. 

%%%%%%%%%%%%%%%%%%%%%%%%%%%%%%%%%%%%%%%%%%%%%%%%%%%%%%%%%%%%%%%%%%%%%%%%%%%%%%%%%%%%%%%%%%%%%
\subsection{Lossy compression case}

\par
In this section we show the results we obtained by using the BP algorithm as an encoder of the scheme.
\par
In the case of lossy compression, the Edwards-Anderson parameter $q$ vanishes 
as discussed in the references \cite{Hosaka2002,Cousseau2008}, 
implying that $| \left\langle \vec{s}_l \right\rangle |^2=0$ (where $\left\langle \ldots \right\rangle$ 
denotes the average with respect to $\vec{y}$ and $\vec{x}$). 
This means that it is not possible to determine the most probable sign of $s^i_l$. 
To avoid this uncertainty we again introduce a particular prior of the form 
\begin{eqnarray}
  q^t_{il} (s^i_l) & = & e^{s^i_l \tanh^{-1} (\gamma m^t_{il})},
\end{eqnarray}
where $0 \leq \gamma < 1$ denotes an amplitude of the inertia term. %[MODIFIED]
Note that $\gamma$ is set by trial and error. %[ADDED]
This method was already successfully applied by Murayama \cite{Murayama2004}. 
\par
The general procedure is as follows (in each case, 
the threshold parameter $k$ is set to the optimal theoretical value, see the reference \cite{Cousseau2008}). 
First, an original message $\vec{y}$ is generated from the distribution (\ref{ydistrib}). 
Then the original message is turned into a codeword $\vec{s}$ using the BP-based algorithms which are shown in Appendix \ref{appendix.B}. 
The codeword is subsequently decoded into $\hat{\vec{y}}$ 
using the proper tree-like multilayer perceptron decoder network. 
The distortion between the decoded message $\hat{\vec{y}}$ 
and the original message $\vec{y}$ is then computed.
\par
We conducted two types of simulations. In the first one, 
the number of hidden units $K$, the size of the codeword $N$, 
and the bias parameters $p$ of the distribution (\ref{ydistrib}) are kept constant. 
The changing parameter is the size of the original message $M$ 
which results in different values for the code rate $R=N/M$. 
For each value of $R$ tested, we perform $100$ runs. 
For each run, we perform $35$ BP iterations 
and the resulted estimated codeword $\vec{s}$ is then decoded into $\hat{\vec{y}}$. 
The distortion between $\hat{\vec{y}}$  and $\vec{y}$ is then computed. 
The code rate is plotted against the mean value of the distortion.
\par
The second type of experiment is exactly the same as in the error correcting case. 
We fix the value of $K,N,M,p$ and generate an original message $\vec{y}$. 
We let run the BP algorithm and get a codeword $\vec{s}$ after $35$ iterations. 
Then we keep the same original message and let run the BP again 
but with different initial values. After $35$ iterations 
we get another codeword $\vec{s}^{\prime}$. We perform the same procedure $30$ times 
and we calculate the average overlap $\frac 1 N \vec{s} \cdot \vec{s}^{\prime}$ 
between all the obtained codewords. Next we generate a new original message $\vec{y}$ 
and do the same procedure for $50$ different original messages. 
We finally plot the obtained average overlap using histograms, 
thus reflecting the distribution of the codeword space.

%%%%%%%%%%%%%%%%%%%%%%%%%%%%%%%%%%%%%%%%%%%%%%%%%%%%%%%%%%%%%%%%%%%%%%%%%%%%%%%%%%%%%%%%%%%%%
\subsubsection{Parity tree with non-monotonic hidden units (PTH)}

\par
\begin{figure}[t]
  \vspace{0mm}%<--space
  \begin{center}
  \includegraphics[width=0.6\linewidth,keepaspectratio]{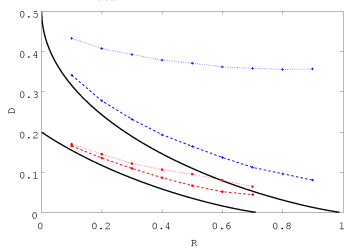}
  \end{center}
  \caption{Empirical performance of the BP-based encoder for lossy compression using the PTH 
    with $K=1$ and $K=3$ for unbiased messages ($p=0.5$, on the top of the figure) 
    and biased message ($p=0.8$, on the bottom). 
    Dashed lines are for $K=1$ and dotted lines are for $K=3$. 
    We used $N=1000$ for $K=1$, and $N=999$ for $K=3$. 
    The inertia term $\gamma=0.45$ was set by trial and error. 
    The solid lines give the Shannon bound. 
    The top one is for $p=0.5$, the bottom one is for $p=0.8$.}
  \label{fig:PTH_Dist}
  \vspace{0mm}%<--space
\end{figure}

We show the results obtained for the PTH with $K=1$ and $K=3$ hidden units 
for unbiased message (i.e.: $p=0.5$) an biased message with $p=0.8$ in Figure \ref{fig:PTH_Dist}. 
The solid line represents the rate distortion function corresponding to the Shannon bound, 
that is the lowest achievable distortion for a given code rate $R$. 
The average distortion for $100$ trials is plotted.
\par
For $K=1$, the results are quite far from the Shannon bound 
for large code rate but approaches it for small ones (for both $p=0.5$ and $p=0.8$). 
We find the same results as in Hosaka et al. \cite{Hosaka2006}. 
Then, the same tendency can be shown for biased messages with $p<0.5$ 
but in those cases, for symmetry reasons, %[MODIFIED] but however -> but
we should use $-f_k$ as a transfer function 
which gives slightly different BP equations (some signs change). 
Hence, for simplicity we restrict the present study to biased messages with $p>0.5$.
\par
The result for unbiased message and $K=3$ are extremely bad 
and the BP does not seems to converge to any relevant codeword. 
This is surprising. 
On top of that, while the performance are also poorer than $K=1$ for biased messages, 
it is not as extreme as for the unbiased case. 
The reasons for such a behavior are not very clear. 
The codeword space structure is again clearly affected 
when more than one hidden unit is used and is likely to perturb the BP dynamics. 
Bias in the original message seems to be another factor to take into account.
\par
\begin{figure}[t]
  \begin{center}
  \small
  \includegraphics[width=0.6\linewidth,keepaspectratio]{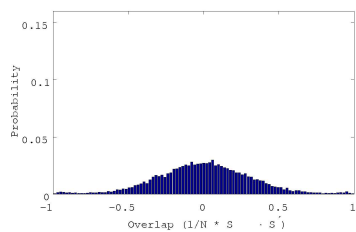} \\
  {\footnotesize (a) $K=1$, $N=100$, $p=0.5$} \\
  \includegraphics[width=0.6\linewidth,keepaspectratio]{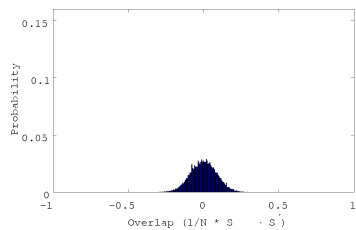} \\
  {\footnotesize (b) $K=1$, $N=1000$, $p=0.5$} \\
  \end{center}
  \caption{Overlap of the solutions given by the BP-based encoder for lossy compression using the PTH 
    with $R=0.4$ and $\gamma=0.45$ which is set by trial and error. 
    The Shannon bound is $0.15$ for $p=0.5$ and $0.057$ for $p=0.8$.
    (a) $K=1$ and $N=100$ and $p=0.5$. The empirical distortion over the trial is $0.21$. 
    (b) $K=1$ and $N=1000$ and $p=0.5$. The empirical distortion is $0.19$. 
  }
  \label{fig:PTH_K1_N100_LCDistrib}
  \vspace{0mm}%<--space
\end{figure}

Then we try to investigate the structure of the codeword space. 
We plot the histograms of the overlap of the codewords obtained 
using the BP for $K=1$, $N=100$, $R=0.4$ and $p=0.5$ 
in Figure \ref{fig:PTH_K1_N100_LCDistrib} (a). 
In this case, it is interesting to note that despite one might believe, 
the BP does not converge to two different solutions. 
As discussed in the reference \cite{Cousseau2008}, for $K=1$, 
we have at least two optimal codewords $\pm \vec{s}$. 
Therefore, one might expect to see two peaks concentrated around $\pm 1$ 
but this is not the case. There are two very small peaks around $\pm 1$ 
and one large peak with its center around $0$. 
This implies that there are many codewords completely uncorrelated 
which share very similar distortion properties. 
To confirm this conjecture, we perform exactly the same experiment 
but with a larger codeword size $N=1000$. Results are shown 
in Figure \ref{fig:PTH_K1_N100_LCDistrib} (b). 
\par
This time, the small peaks around $\pm 1$ completely vanish 
and we have a Gaussian like distribution centered on $0$. 
This confirms the fact that there is a very large amount of uncorrelated codewords 
sharing the same distortion properties. 
This is a surprising result. 
\par
\begin{figure}[t]
  \begin{center}
  \small
  \includegraphics[width=0.6\linewidth,keepaspectratio]{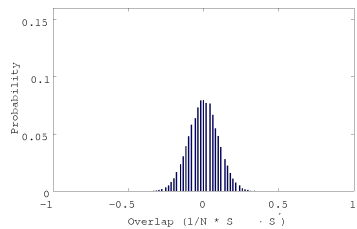} \\
  {\footnotesize (a) $K=3$, $N=102$, $p=0.5$} \\
  \includegraphics[width=0.6\linewidth,keepaspectratio]{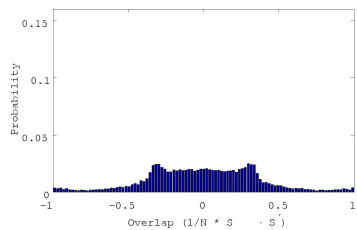} \\
  {\footnotesize (b) $K=3$, $N=102$, $p=0.8$} \\
  \includegraphics[width=0.6\linewidth,keepaspectratio]{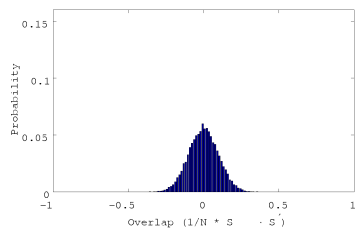} \\
  {\footnotesize (c) $K=3$, $N=999$, $p=0.8$} 
  \end{center}
  \caption{Overlap of the solutions given by the BP-based encoder for lossy compression using the PTH 
    with $R=0.4$ and $\gamma=0.45$ which is set by trial and error. 
    The Shannon bound is $0.15$ for $p=0.5$ and $0.057$ for $p=0.8$.
    (a) $K=3$ and $N=102$ and $p=0.5$. The empirical distortion is $0.43$. 
    (b) $K=3$ and $N=102$ and $p=0.8$. The empirical distortion is $0.118$. 
    (c) $K=3$ and $N=999$ and $p=0.8$. The empirical distortion is $0.101$. 
  }
  \label{fig:PTH_K1_N100_LCDistrib_2}
  \vspace{0mm}%<--space
\end{figure}

We perform the same type of experiment but with $K=3$. 
We first consider unbiased messages ($p=0.5$). 
We show the result for $N=102$ only 
because there is no major change with larger value of $N$ in this case. 
Results are plotted in Figure \ref{fig:PTH_K1_N100_LCDistrib_2} (a). 
We obtain a Gaussian like distribution centered on $0$. 
This means that the solution given by the BP are almost uncorrelated 
between each others. 
This time they do not correspond to any relevant solution 
as indicated by the empirical distortion 
which is close to $0.5$ meaning completely random codewords. 
It seems that for $K>1$ and for unbiased messages ($p=0.5$), 
the number of suboptimal states becomes very large 
and the BP fails to converge to any relevant codeword. 
\par
Then we consider biased messages ($p=0.8$). 
Figure \ref{fig:PTH_K1_N100_LCDistrib_2} (b) shows the results for $N=102$. 
In this case we have two peaks located at $\pm 1/3$ linked 
by a rather high plateau and two small peaks at $\pm 1$. 
The peaks location corresponds to the $2^K$ possible combinations 
of codewords ensured by the structure 
of the network (discussed in the reference \cite{Cousseau2008}). 
This means that in many cases the BP converge to one of this possible $2^K$ combination. 
However the rather high plateau centered on $0$ shows 
that the BP converges many time to uncorrelated codewords. 
This means that on top of the $2^K$ codewords sharing the same distortion properties, 
we have a large number of uncorrelated codewords 
which share rather similar distortion properties. 
We decide to investigate the same case but with a larger value of $N$. 
Figure \ref{fig:PTH_K1_N100_LCDistrib_2} (c) shows the result for $N=999$. 
This time, the peaks completely vanish 
and we obtain a Gaussian like distribution centered on $0$. 
The empirical distortion obtained $0.101$ shows 
that the BP converges to a relevant solution (even if not optimal). 
This shows that as $N$ gets larger, the number of uncorrelated codewords 
sharing similar distortion properties becomes extremely large. 
This is an interesting feature. 
However, the results are not Shannon optimal and as $K$ increases, 
the results for biased messages becomes smoothly worse and worse. 
Nevertheless, the reason why the BP fails to work 
for unbiased messages when $K>1$ is still unclear.
\par
To conclude the case of the PTH, we can say that for $K=1$, 
the BP converges but with relatively poor performance. 
The codeword space exhibits an interesting structure, 
showing that many uncorrelated codewords share very similar properties. 
As the codeword length gets larger, the number of these codewords 
sharing very similar distortion properties seem to increase dramatically. 
For $K>1$, the performance smoothly deteriorates for biased messages 
but for near unbiased ones, the BP fails. 
This is probably due to the rise of suboptimal states when using more than $1$ hidden unit. 
The geometrical structure of the codeword space remains to be investigated. 

%%%%%%%%%%%%%%%%%%%%%%%%%%%%%%%%%%%%%%%%%%%%%%%%%%%%%%%%%%%%%%%%%%%%%%%%%%%%%%%%%%%%%%%%%%%%%
\subsubsection{Committee tree with non-monotonic hidden units (CTH)}

\par
\begin{figure}[t]
  \vspace{0mm}%<--space
  \begin{center}
  \includegraphics[width=0.6\linewidth,keepaspectratio]{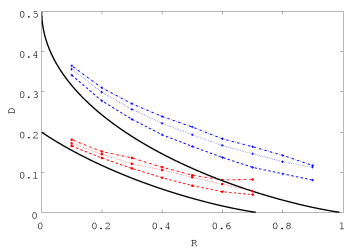}
  \end{center}
  \caption{Empirical performance of the BP-based encoder for lossy compression using the CTH with $K=1$, $K=3$ and $K=5$ 
    for unbiased messages ($p=0.5$, on the top of the figure) 
    and biased message ($p=0.8$, on the bottom). 
    Dashed lines are for $K=1$, dotted lines are for $K=3$ and dash dotted lines 
    are for $K=5$. We used $N=1000$ for $K=1$ and $K=5$, and $N=999$ for $K=3$. 
    The inertia term $\gamma=0.4$ was set by trial and error. 
    The solid lines give the Shannon bound. 
    The top one is for $p=0.5$, the bottom one is for $p=0.8$.
  }
  \label{fig:CTH_Dist}
  \vspace{0mm}%<--space
\end{figure}

We show the results obtained for the CTH 
with $K=1$, $K=3$ and $K=5$ hidden units for unbiased 
and biased messages in Figure \ref{fig:CTH_Dist}. 
We remind that when $K=1$, the CTH is equivalent to the PTH. 
The solid lines represent the rate distortion function 
corresponding to the Shannon bound. 
The average distortion for $100$ trials is plotted. 
\par
The results are quite far from the Shannon bound 
for large code rate but approaches it for small ones. 
However, as $K$ increases, the performance smoothly decreases 
implying that the number of suboptimal states steadily increases 
with the number of hidden units. 
Nevertheless it should be noted that for unbiased messages, 
whereas the BP completely fails in the PTH case for $K>1$, 
this is not the case here. 
Anyway, in the CTH case also, the reason for the deterioration 
of the performance is clearly linked with the number of hidden units. 
\par
\begin{figure}[t]
  \vspace{0mm}%<--space
  \begin{center}
  \small
  \includegraphics[width=0.6\linewidth,keepaspectratio]{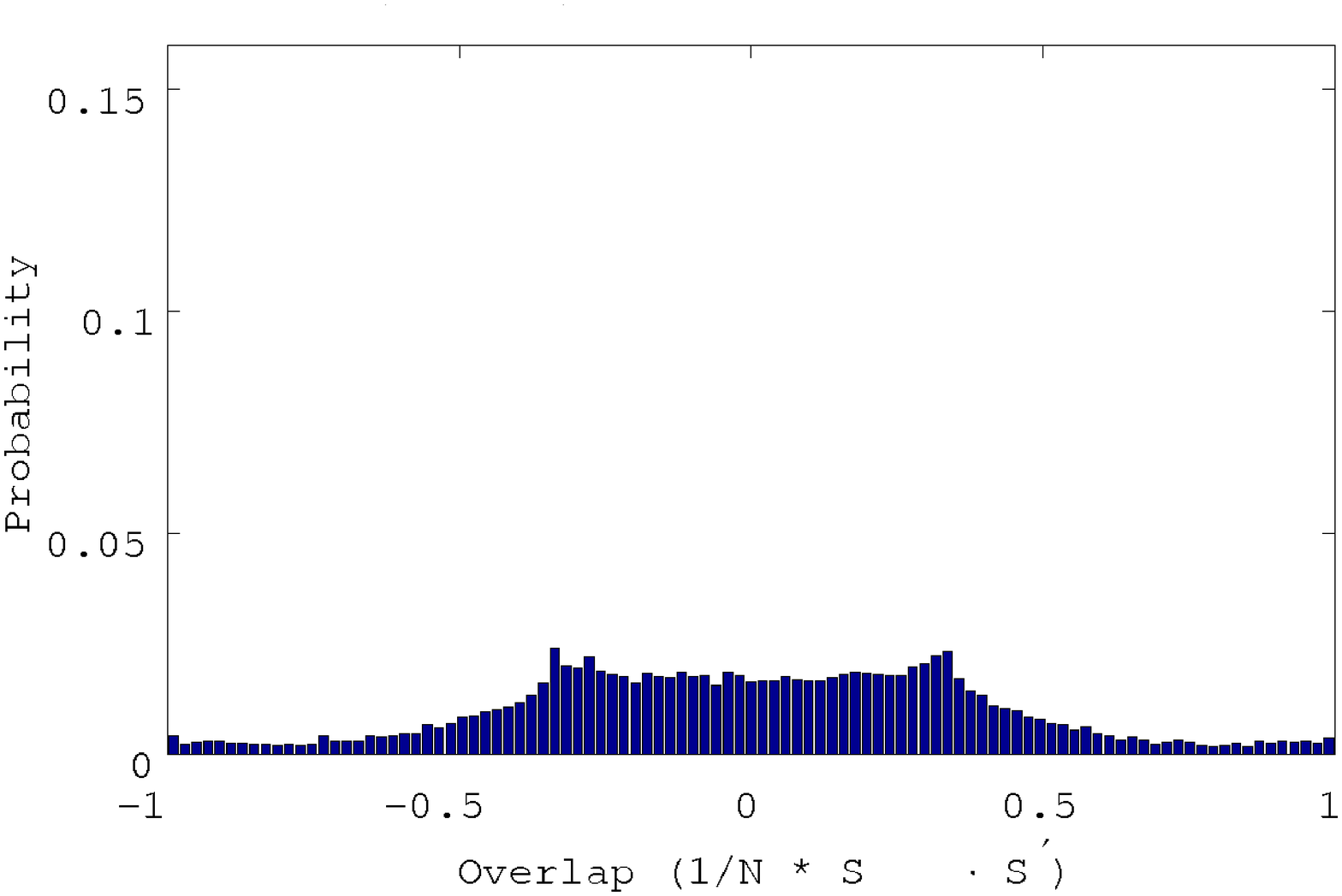} \\
  {\footnotesize (a) $N=102$} \\
  \includegraphics[width=0.6\linewidth,keepaspectratio]{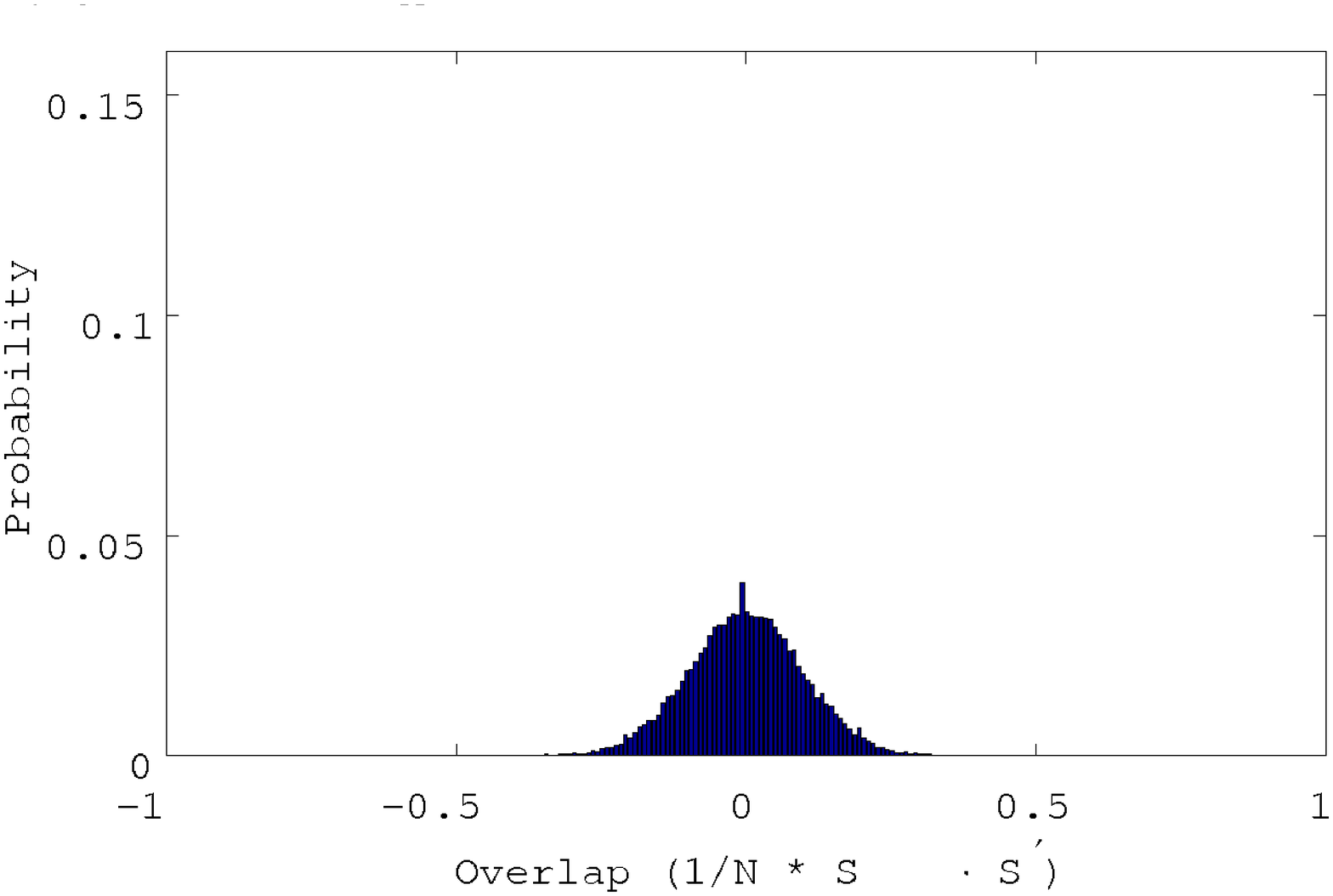} \\
  {\footnotesize (b) $N=1002$} 
  \end{center}
  \caption{Overlap of the solutions given by the BP-based encoder for lossy compresion using the CTH 
    with $K=3$, $R=0.4$, $p=0.8$ and $\gamma=0.4$ which is set by trial and error. 
    The Shannon bound is $0.14$. 
    (a) $N=102$. The empirical distortion is $0.3$. 
    (b) $N=1002$. The empirical distortion is $0.22$.}
  \label{fig:CTH_K3_N100_LCDistrib}
  \vspace{0mm}%<--space
\end{figure}

Then we try to investigate the structure of the codeword space. 
We consider only $K=3$ because $K=1$ is equivalent to the PTH. 
We plot the histograms of the overlap of the solutions obtained 
using the BP in Figure \ref{fig:CTH_K3_N100_LCDistrib} (a). 
In this case, for $K=3$, we have four peaks. 
Two small ones around $\pm 1$ and two big ones linked 
by a plateau around $\pm 1/3$. 
This is the same situation as the PTH with $K=3$, $p=0.8$ and $N=102$. 
The four peaks corresponds to the $2^K$ possible combinations 
of codewords ensured by the structure of the network (discussed in the reference \cite{Cousseau2008}). 
On the other hand, the plateau around $0$ shows 
that there is also many codewords completely uncorrelated 
which share very similar distortion properties. 
To confirm this conjecture, we perform exactly the same experiment 
but with a larger codeword size $N=1002$. 
Results are shown in Figure \ref{fig:CTH_K3_N100_LCDistrib} (b). 
This time, the peaks vanish and we have a Gaussian like distribution centered on $0$. 
This confirm the fact that there is a very large amount of uncorrelated codewords 
sharing the same distortion properties. 
We have the same surprising result as in the PTH case.
\par
To conclude the case of the CTH, we can say that the BP converges but with quite poor performance. 
Furthermore, as $K$ increases, the performance smoothly deteriorates. 
The codeword space exhibits an interesting structure, 
showing that many uncorrelated codewords share very similar distortion properties. 
As the codeword length gets larger, the number of these codewords seems 
to increase dramatically. 
However the reasons of this performance deterioration as $K$ gets larger remains unclear. 
It is likely that the use of several hidden units induces structural change 
in the codeword space and that these are responsible for the BP bad behavior. 

%%%%%%%%%%%%%%%%%%%%%%%%%%%%%%%%%%%%%%%%%%%%%%%%%%%%%%%%%%%%%%%%%%%%%%%%%%%%%%%%%%%%%%%%%%%%%
\subsubsection{Committee tree with a non-monotonic output unit (CTO)}

\par
\begin{figure}[t]
  \vspace{0mm}%<--space
  \begin{center}
  \includegraphics[width=0.6\linewidth,keepaspectratio]{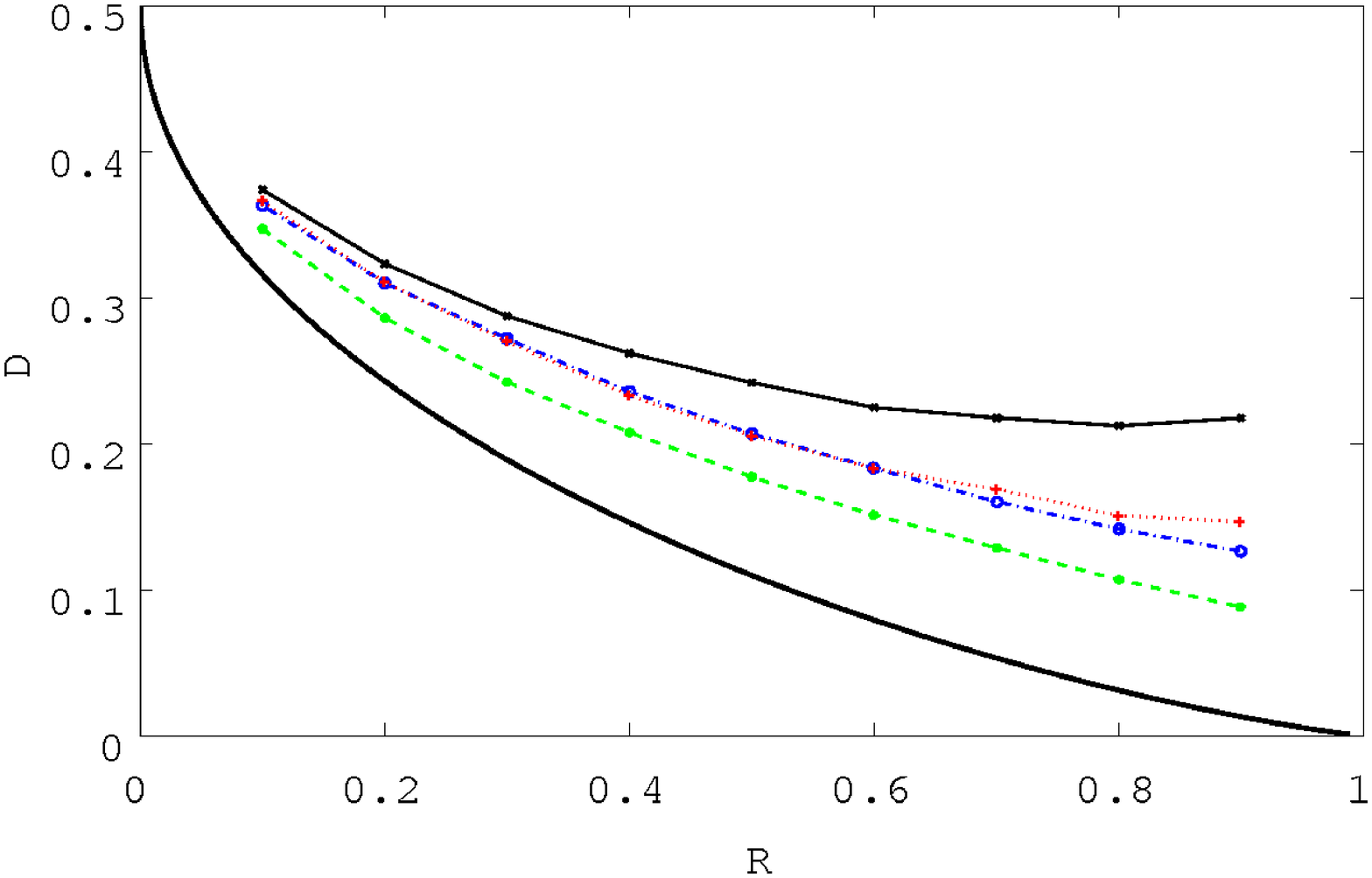} \\
  {\footnotesize (a) $p=0.5$} \\
  \includegraphics[width=0.6\linewidth,keepaspectratio]{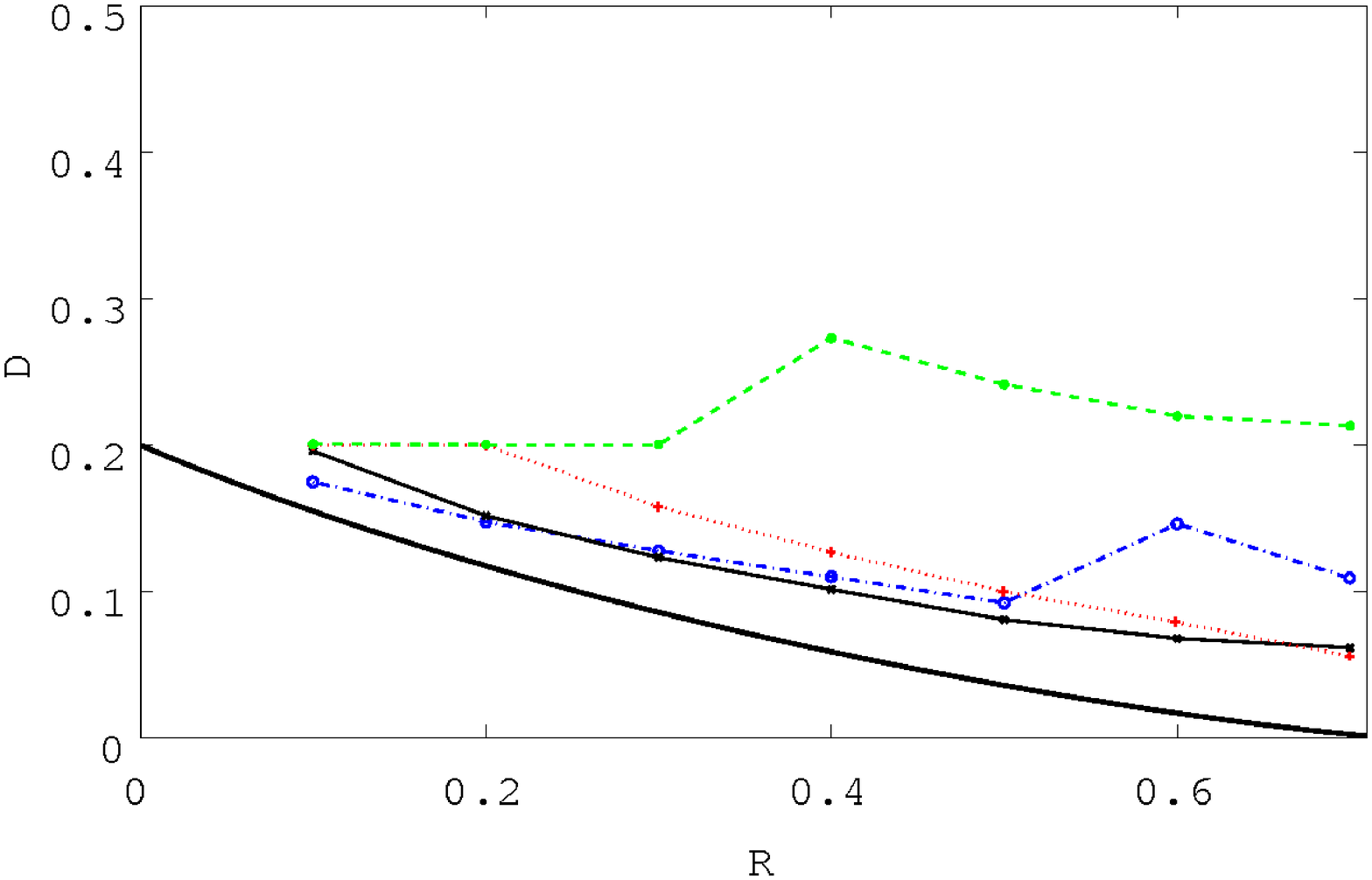} \\
  {\footnotesize (b) $p=0.8$} 
  \end{center}
  \caption{Empirical performance of the BP-based encoder for lossy compression using the CTO 
    with $K=2$, $K=3$, $K=4$ and $K=5$ 
    for unbiased messages ($p=0.5$). 
    Dashed line is for $K=2$, dotted line is for $K=3$, 
    solid line is for $K=4$ and dash dotted line is for $K=5$. 
    We used $N=1000$ for $K=2, 5$, $N=999$ for $K=3$, and $N=1004$ for $K=4$. 
    The inertia term $\gamma=0.4$ was set by trial and error. 
    The continuous solid line (bottom) gives the Shannon bound.
    (a) $p=0.5$. 
    (b) $p=0.8$. 
  }
  \label{fig:CTO_Dist_p05}
  \vspace{0mm}%<--space
\end{figure}

We show the results obtained for the CTO 
with $K=2$, $K=3$, $K=4$ and $K=5$ hidden units for unbiased messages in Figure \ref{fig:CTO_Dist_p05} (a), 
where the CTO cannot be defined for $K=1$. 
The continuous solid line represents the rate distortion function 
corresponding to the Shannon bound. 
The average distortion for $100$ trials is plotted. 
The results for unbiased messages ($p=0.5$) are quite similar as in the CTH case. 
The best performance is obtain for the smaller $K$ and then smoothly deteriorates. 
However, the results do not deteriorate steadily 
(for example $K=5$ gives better performance compared to $K=4$). 
This is probably due to the fact that the free energy is 
a discontinuous function of $K$ as shown in the reference \cite{Cousseau2008}. 
This is then not surprising that the performance 
do not evolve smoothly with $K$. 
On top of that, let us remind that the CTO is expected to give 
the Shannon optimal performance only for an infinite number of hidden units $K$. 
So it is fair the results are quite far from being Shannon optimal. 
However, as $K$ increases, one may expect the performance to become closer 
to the Shannon bound but this is not the case. 
This shows again that a larger number of hidden units clearly 
penalizes the BP performance. 
\par
Next we perform the same experiment but for biased messages with $p=0.8$. 
The results are given in Figure \ref{fig:CTO_Dist_p05} (b). 
The results for biased messages with $p=0.8$ exhibits strange behavior. 
The best performance for small rates $R<0.2$ is obtained for $K=3$ and for $R>0.2$, 
the best performance is given for $K=4$. 
We have some strange jump in performance for $K=2$ between $R=0.3$ and $0.4$ 
and for $K=5$ between $R=0.5$ and $R=0.6$ for example. 
This is probably due to the fact that the tuning of the threshold parameter $k$ 
follows a discontinuous function of $D$ which can explain this kind of discontinuous jump. 
The results are hard to interpret but we observed that for $K>5$, 
the general tendency is to get worse performance. 
As mentioned earlier, the CTO is expected to give Shannon optimal performance 
only for an infinite number of hidden units $K$ so it is fair for the results 
not to be Shannon optimal, especially for small $K$. 
However, as $K$ increases, one may expect the performance to become closer 
to the Shannon bound but this is not the case after $K=4$. 
This shows again that a larger number of hidden units clearly penalizes the BP performance. 
\par
\begin{figure}[t]
  \vspace{0mm}%<--space
  \begin{center}
  \includegraphics[width=0.6\linewidth,keepaspectratio]{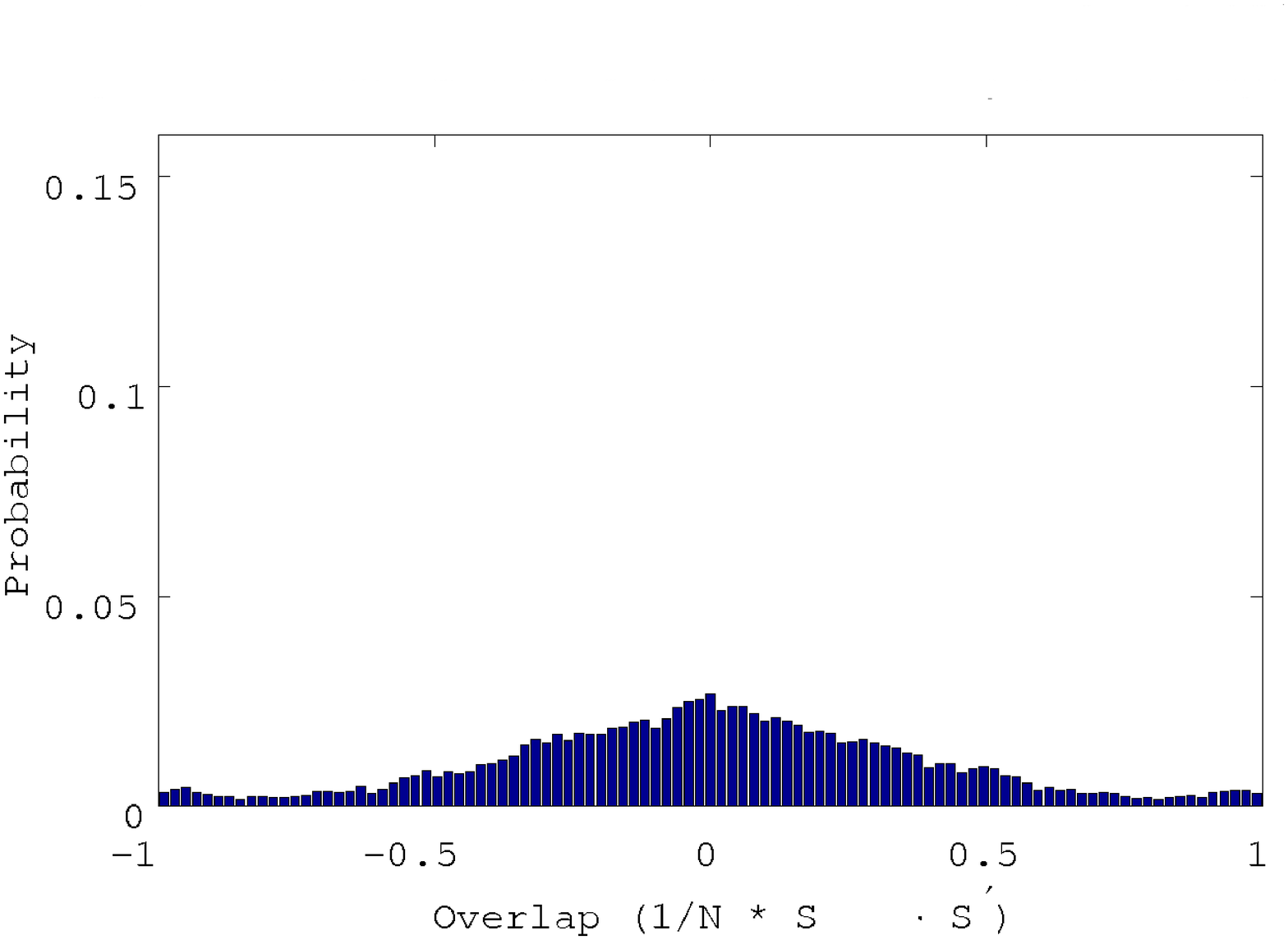} \\
  {\footnotesize (a) $N=100$} \\
  \includegraphics[width=0.6\linewidth,keepaspectratio]{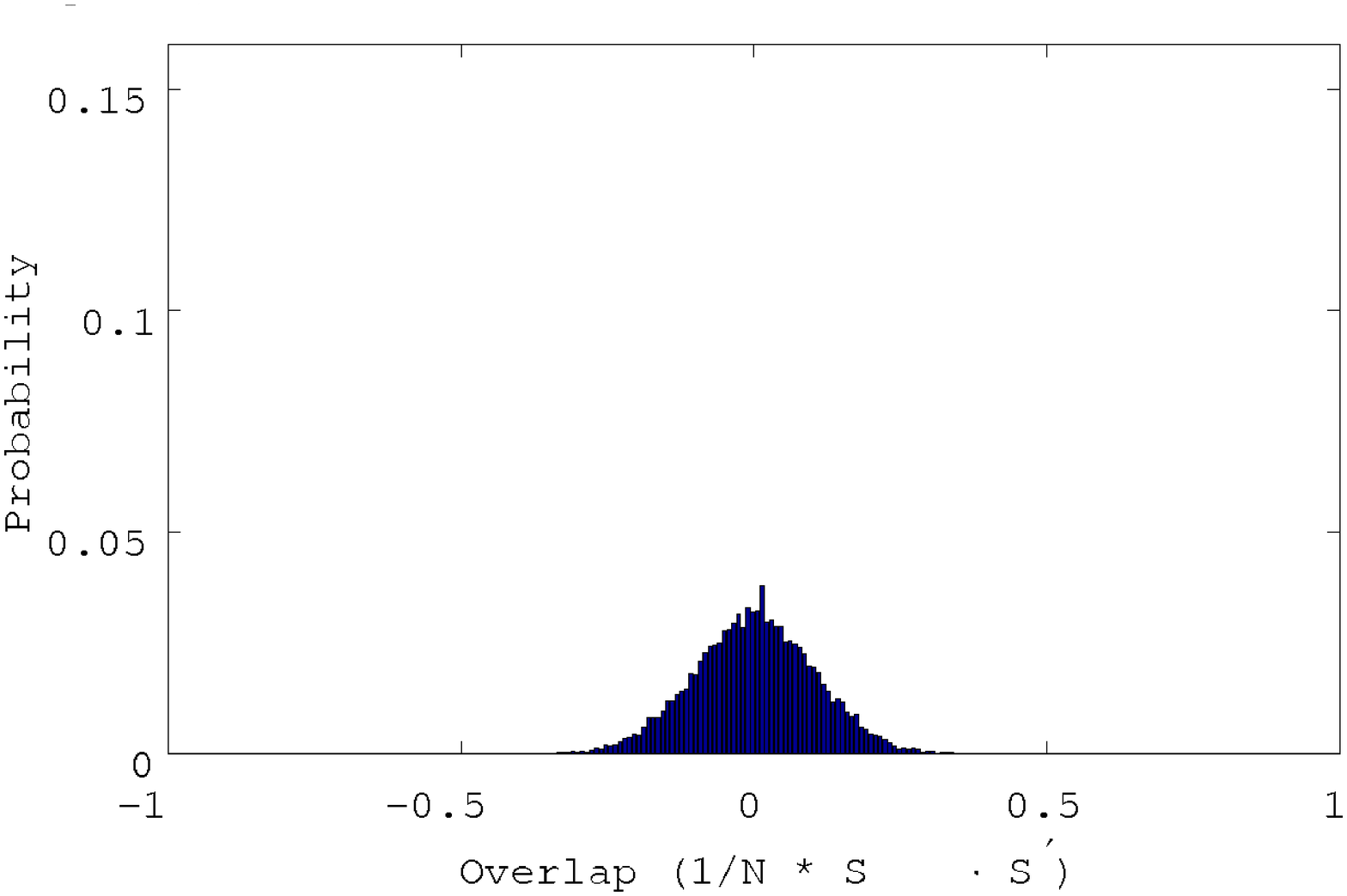} \\
  {\footnotesize (b) $N=1000$} 
  \end{center}
  \caption{Overlap of the solutions given by the BP-based encoder for lossy compression using the CTO 
    with $K=2$, $R=0.4$, $p=0.5$ 
    and $\gamma=0.4$ which is set by trial and error. 
    The Shannon bound is $0.15$.
    (a) $N=100$. The empirical distortion over the trial is $0.25$. 
    (b) $N=1000$. The empirical distortion over the trial is $0.21$. }
  \label{fig:CTO_K2_N100_LCDistrib}
  \vspace{0mm}%<--space
\end{figure}

Then we try to investigate the structure of the codeword space. 
We show the case when $K=2$ only here because the other ones are similar. 
We plot the histograms of the overlap of the solutions obtained 
using the BP for $N=102$ in Figure \ref{fig:CTO_K2_N100_LCDistrib} (a). 
In this case, we have almost the same picture as in the PTH/CTH case with $K=1$. 
Two small peaks around $\pm 1$ and one large plateau around $0$. 
The small peaks corresponds to the codewords 
which share exactly the same distortion properties 
as ensured by the mirror symmetry of the function $f_K$ (discussed in the reference \cite{Cousseau2008}). 
On the other hand, the plateau around $0$ shows 
that there are also many codewords completely uncorrelated 
which share very similar distortion properties. 
To confirm this conjecture, we perform exactly the same experiment 
but with a larger codeword size $N=1000$. Results are shown 
in Figure \ref{fig:CTO_K2_N100_LCDistrib} (b). 
This time, the small peaks around $\pm 1$ completely vanish 
and we have a Gaussian like distribution centered on $0$. 
This confirm the fact that there is a very large amount 
of uncorrelated codewords sharing the same distortion properties.
\par
To conclude the case of the CTO, we can say that the BP converges 
but with quite poor performance. 
On top of that, because of the discontinuous free energy, 
we observe some strange behavior like sudden jump in performance. 
The CTO theoretically gives Shannon performance 
for an infinite number of hidden units $K$ so one may expect the performance given 
by the BP to get better and better as $K$ increases however this is not the case. 
For $K>4$, we generally get poorer and poorer performance showing one more time 
that there is some intimate link between the BP performance and the number of hidden units. 
Finally, as already found for the PTH and CTH, as the codeword length gets larger, 
the number of codewords sharing similar distortion properties seems to increase dramatically. 
The geometrical feature of the codeword space remains to be investigated. 

%%%%%%%%%%%%%%%%%%%%%%%%%%%%%%%%%%%%%%%%%%%%%%%%%%%%%%%%%%%%%%%%%%%%%%%%%%%%%%%%%%%%%%%%%%%%%
\section{Conclusion and Discussion}
\label{sec:discussion}

\par
We have investigated the BP algorithm as a decoder of an error correcting code scheme 
based on tree-like multilayer perceptron encoder. 
In the same way, we have investigated the BP algorithm as a potential encoder of a lossy compression scheme 
based on tree-like multilayer perceptron decoder. 
We have discussed that whether the BP can give practical algorithms or not in these schemes. %[ADDED]
Unfortunately, the BP implementations in those kind of fully connected networks shows strong limitation, 
while the theoretical results seems a bit promising. %[ADDED]
Instead, it reveals it might have a rich and complex structure of the solution space via the BP-based algorithms. %[ADDED]
\par
While these two schemes have been shown to yield the Shannon optimal performance theoretically 
(under some specific conditions, Cf. the references \cite{Cousseau2008bis, Cousseau2008}), 
they lack a practical formal decoder and encoder, respectively. 
The BP algorithm has been proposed as a way to calculate the marginalized posterior probabilities 
of the relevant Boltzmann factor but exhibits poor performance preventing this kind of schemes from being practical. 
The number of hidden units should be kept as small as possible as no gain have been observed by using several ones. 
While the precise reasons behind this bad behavior are still unclear at the present time, 
there is no doubt that the number of hidden units have some deep impact 
onto the solution space of the considered network, which is infered from behavior of the BP-based algorithms. 
It is very probable that the existence of mirror symmetry in the network 
is at the origin of the BP failure. 
It is also very likely that a singular structure similar 
to the one studied in the first part of this paper, 
prevents the standard BP algorithm to work efficiently. 
This underline the necessity to investigate the geometrical feature 
of the solution space of the PTH/CTH/CTO as well as the BP dynamics 
to understand why the BP does not work well when a large $K$ is used. 
It would be interesting to investigate the information geometrical counterpart 
of the BP algorithm to see how well it can performed. 
This remains a future topic of research. 
\par
On the other hand, as discussed in the reference \cite{Hosaka2002} and in the reference \cite{Cousseau2008}, 
the mirror symmetry seems to be a key factor to achieve Shannon performance 
while using perceptron like network in the lossy compression case. 
Since we have $f_k(\vec{s})=f_k(-\vec{s})$, 
one would expect to get two optimal solutions (when $K=1$) 
or more (due to the possible combinations of $\pm \vec{s}_l$) 
but this is not the case. Using a small value of $N$, 
the expected peaks induced by the structure of the network are indeed observable 
but a large concentration of uncorrelated codewords is also visible. 
Using a sufficiently large $N$, those peaks completely vanish, 
and one will always get uncorrelated codewords, trial after trial, 
demonstrating that a very large amount of uncorrelated codewords 
share very similar distortion properties. 
The origin of such particular space structure remains unclear. 
In the same way, the complete failure of the BP 
in the case of the PTH with $K>1$ remains to be investigated. 
We might be able to investigate such problems by evaluating the complexity \cite{Obuchi2009} of the systems. 
This is a part of our future works.

%%%%%%%%%%%%%%%%%%%%%%%%%%%%%%%%%%%%%%%%%%%%%%%%%%%%%%%%%%%%%%%%%%%%%%%%%%%%%%%%%%%%%%%%%%%%%
%%  ACKNOWLEDGEMENTS
\section*{Acknowledgments}

The authors would like to thank Tadaaki Hosaka for his valuable discussions. 
This work was partially supported by a Grant-in-Aid for Encouragement of
Young Scientists (B) (Grant No. 18700230), Grant-in-Aid for Scientific
Research on Priority Areas (Grant Nos. 18079003, 18020007),
Grant-in-Aid for Scientific Research (C) (Grant No. 16500093), and a
Grant-in-Aid for JSPS Fellows (Grant No. 06J06774) from the Ministry of
Education, Culture, Sports, Science and Technology of Japan.

%%%%%%%%%%%%%%%%%%%%%%%%%%%%%%%%%%%%%%%%%%%%%%%%%%%%%%%%%%%%%%%%%%%%%%%%%%%%%%%%%%%%%%%%%%%%%
%%  APPENDIX
\vspace*{5mm}
\appendix

%%%%%%%%%%%%%%%%%%%%%%%%%%%%%%%%%%%%%%%%%%%%%%%%%%%%%%%%%%%%%%%%%%%%%%%%%%%%%%%%%%%%%%%%%%%%%
%%  APPENDIX A
\section{Derivation of the BP decoder for error correcting code case \label{appendix.A}}
\par
The $s^i_l$ are Ising variables, we can reparameterize the above probabilities 
using their corresponding expectation values for the random variable $s^i_l$,
\begin{eqnarray}
  & &
  \hat{\rho}^t_{\mu i l} (s^i_l) 
  = \frac {1+ \hat{m}^t_{\mu i l} s^i_l} 2 , \label{mhat}
  \\
  & &
  \rho^t_{\mu i l} (s^i_l) 
  = \frac {1+ m^t_{\mu i l} s^i_l} 2 , \label{BPm}
  \\
  & &
  p^t (s^i_l | \vec{y} , \{ \vec{x} \} ; \beta) 
  = \frac {1+ m^t_{i l} s^i_l} 2 , \label{mMarg}
\end{eqnarray}
where $\hat{m}^t_{\mu i l}, m^t_{\mu i l},m^t_{ i l}$ denotes the relevant expectation values at time step $t$. 
Computing the expectation is easier than computing the message itself.
\par
Using the following identity $\ln \frac {1+x} {1-x} =2 \tanh^{-1} x$, 
independently of the scheme and network considered, one can already easily derived the following set of equations,
\begin{eqnarray}
  & & \hspace*{-5mm}
  m^{t+1}_{\mu i l} 
  = \tanh \biggl[ \sum_{\mu^{\prime} \neq \mu}^M \tanh^{-1} \hat{m}^t_{\mu^{\prime} i l} 
    + \frac 1 2 \ln \frac {q^t_{i l} (1)} {q^t_{i l} (-1)} \biggr], 
  \label{BP:m} 
  \\
  & & \hspace{-5mm}
  m^{t+1}_{i l} 
  = \tanh \biggl[ \sum_{\mu=1}^M \tanh^{-1} \hat{m}^t_{\mu i l} 
    + \frac 1 2 \ln \frac {q^t_{i l} (1)} {q^t_{i l} (-1)} \biggr]. 
  \label{BP:mMarg}
\end{eqnarray}
In the error correcting code case, $G_{k,\mu}$ is given by 
\begin{eqnarray}
  & & 
  G_{k,\mu} \biggl( \biggl\{ \sqrt{\frac {K} {N}} \vec{s}_l \cdot \vec{x}^{\mu}_l \biggr\} \biggr) 
  \nonumber \\
  & & 
  = 
  2^{-N/M} \biggl\{ \frac 1 2 
  + \frac {y^{\mu}} 2 [(1-r-p) \mathcal{F}_k \biggl( \biggl\{ 
    \sqrt{\frac K N} \vec{s}_l \cdot \vec{x}_l^{\mu} \biggr\} \biggr) \nonumber \\
  & &
  \qquad 
  + (r-p)] \biggr\} . 
\end{eqnarray}
Note that we put $\beta=1$.

%%%%%%%%%%%%%%%%%%%%%%%%%%%%%%%%%%%%%%%%%%%%%%%%%%%%%%%%%%%%%%%%%%%%%%%%%%%%%%%%%%%%%%%%%%%%%
\subsection{Parity tree with non-monotonic hidden units (PTH)}

In the case of the PTH $\mathcal{F}_k$ is given by, 
\begin{equation}
  \mathcal{F}_k \biggl( \biggl\{ \sqrt{\frac K N} \vec{s}_l \cdot \vec{x}_l^{\mu} \biggr\} \biggr) = 
  \prod_{l=1}^{K} f_k \biggl( \sqrt{\frac K N} \vec{s}_l \cdot \vec{x}_l^{\mu} \biggr) .
\end{equation}
Applying the Taylor expansion, this can be rewritten as 
\begin{equation}
  \mathcal{F}_k \biggl( \biggl\{ \sqrt{\frac K N} \vec{s}_l \cdot \vec{x}_l^{\mu} \biggr\} \biggr) \approx
  f_k \left( \lambda^{\mu}_{i l} + \wedge^{\mu}_{i l} \right)
  \prod_{l^{\prime} \neq l}^{K} f_k \left( \wedge^{\mu}_{i l^{\prime}} \right) ,
\end{equation}
where
\begin{eqnarray}
  \lambda^{\mu}_{i l} = \sqrt{ \frac K N } s^i_l x^{\mu}_{i l}, \quad 
  \wedge^{\mu}_{i l} = \sum_{i^{\prime} \neq i}^{N/K} \sqrt{ \frac K N } s^{i^{\prime}}_l x^{\mu}_{i^{\prime} l}
\end{eqnarray}
and we have neglected the remaining $\{ \lambda^{\mu}_{i,l^{\prime}} | l^{\prime} \neq l \}$ 
of order $O(1/ \sqrt{N})$. 
Note that this approximation is justified by the fact that we suppose $N \to \infty$. 
For the same reason we apply the central limit theorem on the $\wedge^{\mu}_{il}$ and find, 
\begin{eqnarray}
  \wedge^{\mu}_{i l} 
  \sim \mathcal{N} (\bar{\wedge}^{t}_{\mu i l},1-q^{t}_{\mu i l}), 
\end{eqnarray}
where 
\begin{eqnarray}
  \bar{\wedge}^{t}_{\mu i l} 
  = \sqrt{\frac K N} \sum_{i^{\prime} \neq i}^{N/K} m^t_{\mu i^{\prime} l} x^{\mu}_{i^{\prime} l}, \quad 
  q^{t}_{\mu i l} 
  = \frac K N \sum_{i^{\prime} \neq i}^{N/K} (m^t_{\mu i^{\prime} l})^2 .
\end{eqnarray}
We finally get 
\begin{equation}
  \hat{\rho}^t_{\mu i l} (s^i_l) \approx 2^{-N/M} \mathfrak{F}^t_{k,\mu i l} (\lambda^{\mu}_{i l}) 
\end{equation}
where 
\begin{eqnarray}
  & &
  \mathfrak{F}^t_{k,\mu i l} (\lambda^{\mu}_{i l}) 
  =  \int_{-\infty}^{+\infty} \prod_{l=1}^K Dz_l \times
     \Bigg\{ \frac 1 2 + \frac {y^{\mu}} 2 (1-r-p) 
     \nonumber \\ 
  & & 
  \qquad \qquad \qquad 
  \times f_k \left( \lambda^{\mu}_{i l} + \bar{\wedge}^{t}_{\mu i l} 
      + z_l \sqrt{1-q^{t}_{\mu i l}} \right) \nonumber \\
  & &
  \qquad \qquad \qquad 
  \times \prod_{l^{\prime} \neq l}^{K} 
      f_k \left( \bar{\wedge}^{t}_{\mu i l^{\prime}} 
      + z_{l^{\prime}} \sqrt{1-q^{t}_{\mu i l^{\prime}}} \right)
      \nonumber \\ 
  & & 
  \qquad \qquad \qquad 
  + \frac {y^{\mu}} 2 (r-p) \Bigg\} ,
\end{eqnarray}
and 
\begin{eqnarray}
  Dx = \frac {e^{- \frac {x^2} 2} dx} {\sqrt{2 \pi}} .
\end{eqnarray}
Using the fact that $\lambda^{\mu}_{i l}$ is of order $O(1/\sqrt{N})$, 
we expand $\mathfrak{F}^t_{k,\mu i l}$ around $0$ and get, 
\begin{eqnarray}
  \hat{\rho}^t_{\mu i l} (s^i_l) \approx 2^{-\frac NM}
  \biggl( \mathfrak{F}^t_{k,\mu i l} (0) + \lambda^{\mu}_{i l} 
  \left. \frac {\partial \mathfrak{F}^t_{k,\mu i l} (\lambda^{\mu}_{i l})} {\partial \lambda^{\mu}_{i l}} \right|_{\lambda^{\mu}_{i l } = 0} \biggr). 
\end{eqnarray}
Finally, using (\ref{mhat}) we get the expresion of $\hat{m}^t_{\mu il}$ as follows: 
\begin{eqnarray}
  \hat{m}^t_{\mu il} 
  = \sqrt{\frac K N} x^{\mu}_{i l} \frac {\left. 
    \frac {\partial \mathfrak{F}^t_{k,\mu i l} (\lambda^{\mu}_{i l})}{\partial \lambda^{\mu}_{i l}} 
    \right|_{\lambda^{\mu}_{i l } = 0}} 
    {\mathfrak{F}^t_{k,\mu i l} (0)}. 
  \label{BP:mHat}
\end{eqnarray}
Evaluating $\mathfrak{F}^t_{k,\mu i l} (0)$ 
and $\frac {\partial \mathfrak{F}^t_{k,\mu i l} (\lambda^{\mu}_{i l})} {\partial \lambda^{\mu}_{i l}} |_{\lambda^{\mu}_{i l } = 0}$, 
we can explicitly obtain $\hat{m}^t_{\mu il}$. 
\par
So using (\ref{BP:m}), (\ref{BP:mMarg}) and (\ref{BP:mHat}) iteratively until a fixed point is reached, 
one should be able to decode the received corrupted codeword $\vec{y}$ 
and find back the original message $\vec{s}^0$. 
However, this procedure still requires $O(N^3)$ operations 
so one might want to reduce the complexity of the algorithm. 
\par
For simplicity, we suppose a uniform prior $q^t_{il}$ hereafter. 
However, the results can be easily generalized for more complex priors. 
Since $\hat{m}^t_{\mu i l}$ is of order $O( 1  / \sqrt{N})$, we have 
\begin{eqnarray}
  m^{t+1}_{\mu i l} 
  &=& \tanh \bigg[ \sum_{\mu^{\prime} \neq \mu}^M \tanh^{-1} \hat{m}^t_{\mu^{\prime} i l} \bigg], \nonumber \\
  &\approx&  m^{t+1}_{i l} - \left[ 1 - (m^{t+1}_{i l})^2 \right] \hat{m}^t_{\mu i l}. 
\end{eqnarray}
We can then evaluate the following equations using the above approximation, 
\begin{eqnarray}
  q^t_{\mu i l} 
  = \frac K N \sum_{i^{\prime} \neq i}^{N/K} (m^t_{\mu i^{\prime} l})^2 
  \approx q^t_l - \hat{q}^t_{\mu l} - (\varepsilon^t_{il})^2. 
\end{eqnarray}
where 
\begin{eqnarray}
  & & q^t_l = \frac K N \sum_{i=1}^{K/N} (m^t_{il})^2, \\
  & & \hat{q}^t_{\mu l} = 2 \frac K N \sum_{i=1}^{N/K} m^t_{i l}
      \left( 1-[m^t_{i l}]^2 \right) \hat{m}^{t-1}_{\mu i l}, \\
  & & \varepsilon^t_{il} = \sqrt{\frac K N} m^t_{i l}.
\end{eqnarray}
We here insert the lacking term in the partial sum ($\sum_{i^{\prime} \neq i} \approx \sum_{i=1}$) 
of the cross term since this should be negligible for large $N$. 
In the same way we have 
\begin{eqnarray}
  \bar{\wedge}^t_{\mu i l} 
  = \sqrt{\frac K N} \sum_{i^{\prime} \neq i}^{N/K} m^t_{\mu i^{\prime} l} x^{\mu}_{i^{\prime} l} 
  \approx \bar{\wedge}^t_{\mu l} - \hat{\wedge}^t_{\mu l} - x^{\mu}_{il} \varepsilon^t_{i l}, 
\end{eqnarray}
where
\begin{eqnarray}
  & & \bar{\wedge}^t_{\mu l} = \sqrt{\frac K N} \sum_{i=1}^{K/N} m^t_{il} x^{\mu}_{il}, \\
  & & \hat{\wedge}^t_{\mu l} = \sqrt{\frac K N} \sum_{i=1}^{N/K} 
      \left( 1-[m^t_{i l}]^2 \right) \hat{m}^{t-1}_{\mu i l} x^{\mu}_{il}.
\end{eqnarray}
\par
Using these equations, we can rewrite $\mathfrak{F}^t_{k,\mu il}$ and its derivative as a function of $\varepsilon^t_{il}$,
\begin{eqnarray}
  & & \left. \frac {\partial \mathfrak{F}^t_{k,\mu i l} (\lambda^{\mu}_{i l})} {\partial \lambda^{\mu}_{i l}} \right|_{\lambda^{\mu}_{i l } = 0} 
      \equiv \hat{U}^t_{k,\mu il} (\varepsilon^t_{il}), \\
  & & \mathfrak{F}^t_{k,\mu i l} (0) 
      \equiv \hat{V}^t_{k,\mu il} (\varepsilon^t_{il}),
\end{eqnarray}
Then, because each $\varepsilon^t_{il}$ is of order $O (1/\sqrt{N})$, 
we approximate $\{ \hat{U}^t_{k,\mu il} , \hat{V}^t_{k,\mu il} \}$ by $\{ U^t_{k,\mu il} , V^t_{k,\mu il} \}$ 
where we neglect all the terms $\{ \varepsilon^t_{i l^{\prime}} | l^{\prime} \neq l \}$, which gives 
\begin{eqnarray}
  & & \hat{U}^t_{k,\mu il} (\varepsilon^t_{il}) \approx U^t_{k,\mu il} (\varepsilon^t_{il}), \\
  & & \hat{V}^t_{k,\mu il} (\varepsilon^t_{il}) \approx V^t_{k,\mu il} (\varepsilon^t_{il}). 
\end{eqnarray}
Using this approximation, we get 
\begin{eqnarray}
  \hat{m}^t_{\mu il} \approx \sqrt{\frac K N} x^{\mu}_{il} \Phi^t_{k,\mu i l} (\varepsilon^t_{il}), 
\end{eqnarray}
where we put 
\begin{eqnarray}
  \Phi^t_{k,\mu i l} (\varepsilon^t_{il}) = \frac {U^t_{k,\mu il} (\varepsilon^t_{il})} {V^t_{k,\mu il} (\varepsilon^t_{il})}.
\end{eqnarray}
Then once again, because $\varepsilon^t_{il}$ is of order $O(1/\sqrt{N})$, we perform the Taylor epxansion: 
\begin{eqnarray}
  \hat{m}^t_{\mu il} 
  \approx \sqrt{\frac K N} x^{\mu}_{il} \left[ \Phi^t_{k,\mu i l} (0) 
  + \sqrt{\frac K N} m^t_{il} \left. 
  \frac {\partial \Phi^t_{k,\mu il} (\varepsilon^t_{il})} {\partial \varepsilon^t_{il}} \right|_{\varepsilon^t_{il} = 0} \right]. 
\end{eqnarray}
For simplicity, we hereafter use the following abbreviations: 
\begin{eqnarray}
  & & U^t_{k,\mu il} (0) \equiv U^t_{k,\mu l}, \\
  & & V^t_{k,\mu il} (0) \equiv V^t_{k,\mu l}, \\
  & & \Phi^t_{k,\mu i l} (0) \equiv \Phi^t_{k,\mu l}, \\
  & & \left. \frac{ \partial U^t_{k,\mu il} (\varepsilon^t_{il}) } { \partial \varepsilon^t_{il}}  \right|_{\varepsilon^t_{il}=0} \equiv x^{\mu}_{il} \tilde{U}^t_{k,\mu l}, \\
  & & \left. \frac{ \partial V^t_{k,\mu il} (\varepsilon^t_{il}) } { \partial \varepsilon^t_{il}}  \right|_{\varepsilon^t_{il}=0} \equiv x^{\mu}_{il} \tilde{V}^t_{k,\mu l}, 
\end{eqnarray}
which appear in $\frac {\partial \Phi^t_{k,\mu il} (\varepsilon^t_{il})} {\partial \varepsilon^t_{il}} |_{\varepsilon^t_{il} = 0}$. 
Note that because we neglect all the $\varepsilon^t_{i l^{\prime}}$ and use the value of the above functions evaluated at $0$ only, we can drop the index $i$. 
\par
So using all this results we have (we suppose a uniform prior for simplicity), 
\begin{eqnarray}
  m^{t+1}_{il} 
  & = & \tanh \left[ \sum_{\mu =1}^M \tanh^{-1} (\hat{m}^t_{\mu il}) \right], \nonumber \\
  & = & \tanh \left[ \sum_{\mu =1}^M \sqrt{\frac K N} x^{\mu}_{il} \Phi^t_{k,\mu l} + m^t_{il} \mathfrak{G}^t_{k, il}  \right], 
\end{eqnarray}
where we put 
\begin{eqnarray}
  \mathfrak{G}^t_{k, il} \equiv \frac K N \sum_{\mu=1}^M x^{\mu}_{il} 
  \left. \frac {\partial \Phi^t_{k,\mu il} (\varepsilon^t_{il})} {\partial \varepsilon^t_{il}} \right|_{\varepsilon^t_{il} = 0} .
\end{eqnarray}
Neglecting small order terms, we obtain 
\begin{eqnarray}
  \mathfrak{G}^t_{k, il} 
  = \frac K N \sum_{\mu=1}^M \frac { \tilde{U}^t_{k,\mu l} V^t_{k,\mu l} - \tilde{V}^t_{k,\mu l} U^t_{k,\mu l} } { (V^t_{k,\mu l})^2} 
  \equiv \mathfrak{G}^t_{k,l} .
\end{eqnarray}
We therefore obtain the approximated BP equation as follows: 
\begin{eqnarray}
  m^{t+1}_{il} 
  = \tanh \bigg[ \sum_{\mu=1}^M \sqrt{\frac K N} x^{\mu}_{il} \Phi^t_{k,\mu l}
  + m^t_{il} \mathfrak{G}^t_{k,l} + \frac 1 2 \ln \frac {q^t_{i l} (1)} {q^t_{i l} (-1)} \bigg] , \label{refineBP}
\end{eqnarray}
where we have inserted back the term depending on the prior. 
We then arrive at (\ref{refineBP-text}). 
The BP algorithm is thus finally reduced to (\ref{refineBP}) and requires about $O(N^2)$ operations for each step. 
The MPM estimator at time step $t$ is given by $s^i_l = \sgn (m^{t}_{il})$. 
\par
In this case the BP reduces to a single recurrent equation given by (\ref{refineBP}), 
where in the case of the PTH, we have 
\begin{eqnarray}
  & &
  U^t_{k,\mu l} 
  = \frac {y^{\mu} (1-r-p)} {\sqrt{2 \pi (1-q^t_l)}} 
  \left[ e^{\frac {-1} 2 (w^{t+}_{k,\mu l})^2 } - e^{\frac {-1} 2 (w^{t-}_{k,\mu l})^2 } \right]
  \nonumber \\ 
  & &
  \qquad \qquad 
  \times \prod_{l^{\prime} \neq l}^K 
  \left[ 1 - 2 H (w^{t+}_{k,\mu l^{\prime}}) -2 H (w^{t-}_{k,\mu l^{\prime}}) \right], \\
  & &
  V^t_{k,\mu l} 
  = \frac 1 2 + \frac {y^{\mu}} 2 (r-p) + \frac {y^{\mu}} 2 (1-r-p) \nonumber \\ 
  & & 
  \qquad \qquad 
  \times \prod_{l=1}^K \left[ 1 - 2 H (w^{t+}_{k,\mu l}) -2 H (w^{t-}_{k,\mu l}) \right], \\
  & &
  \tilde{U}^t_{k,\mu l} 
  = \frac {\left[ w^{t+}_{k,\mu l} e^{\frac {-1} 2 (w^{t+}_{k,\mu l})^2 } 
  + w^{t-}_{k,\mu l} e^{\frac {-1} 2 (w^{t-}_{k,\mu l})^2 } \right] U^t_{k,\mu l} } 
  {\sqrt{1-q^t_l} \left[ e^{\frac {-1} 2 (w^{t+}_{k,\mu l})^2 } - e^{\frac {-1} 2 (w^{t-}_{k,\mu l})^2 } \right]} , \\
  & &
  \tilde{V}^t_{k,\mu l} 
  = - U^t_{k,\mu l},
\end{eqnarray}
where 
\begin{eqnarray}
  w^{t+}_{k,\mu l} = \frac {k + \bar{\wedge}^t_{\mu l} - \hat{\wedge}^t_{\mu l} } {\sqrt{1-q^t_l}}, \quad 
  w^{t-}_{k,\mu l} = \frac {k - \bar{\wedge}^t_{\mu l} + \hat{\wedge}^t_{\mu l} } {\sqrt{1-q^t_l}}
\end{eqnarray}
and
\begin{eqnarray}
  H(u) = \int_{u}^{+\infty} Dx . 
\end{eqnarray}
In another schemes, we first calculate 
$U^t_{k,\mu l}$, 
$V^t_{k,\mu l}$, 
$\tilde{U}^t_{k,\mu l}$ and 
$\tilde{V}^t_{k,\mu l}$ 
which are needed to obtain an iterative equation of (\ref{refineBP}).

%%%%%%%%%%%%%%%%%%%%%%%%%%%%%%%%%%%%%%%%%%%%%%%%%%%%%%%%%%%%%%%%%%%%%%%%%%%%%%%%%%%%%%%%%%%%%
\subsection{Committee tree with non-monotonic hidden units (CTH)}

\par
In the case of the CTH, $\mathcal{F}_k$ is given by, 
\begin{eqnarray}
  & & 
  \mathcal{F}_k \biggl( \biggl\{ \sqrt{\frac K N} \vec{s}_l \cdot \vec{x}_l^{\mu} \biggr\} \biggr) \nonumber \\
  & & 
  = \sgn \biggl[ \sum_{l=1}^{K} f_k \biggl( \sqrt{\frac K N} \vec{s}_l \cdot \vec{x}_l^{\mu} \biggr) \biggr] \nonumber \\
  & &
  \approx \sgn \biggl[ f_k \biggl( \lambda^{\mu}_{i l} + \wedge^{\mu}_{i l} \biggr) 
            + \sum_{l^{\prime} \neq l}^{K} f_k \biggl( \wedge^{\mu}_{i l^{\prime}} \biggr) \biggr].
\end{eqnarray}
In the same way as the PTH, we find 
\begin{eqnarray}
  \mathfrak{F}^t_{k,\mu i l} (\lambda^{\mu}_{i l}) 
  &=& \int_{-\infty}^{+\infty} \prod_{l=1}^K Dz_l \times
      \Bigg\{ \frac 1 2 + \frac {y^{\mu}} 2 (1-r-p) \nonumber \\ 
  & & \times \sgn \biggl[ f_k \biggl( \lambda^{\mu}_{i l} 
      + \bar{\wedge}^{t}_{\mu i l} + z_l \sqrt{1-q^{t}_{\mu i l}} \biggr) \nonumber \\
  & & + \sum_{l^{\prime} \neq l}^{K} f_k \biggl( \bar{\wedge}^{t}_{\mu i l^{\prime}} 
      + z_{l^{\prime}} \sqrt{1-q^{t}_{\mu i l^{\prime}}} \biggr) \biggr] \nonumber \\ 
  & & + \frac {y^{\mu}} 2 (r-p) \Bigg\}. 
\end{eqnarray}
Evaluating $\mathfrak{F}^t_{k,\mu i l} (0)$ 
and $\frac {\partial \mathfrak{F}^t_{k,\mu i l} (\lambda^{\mu}_{i l})} {\partial \lambda^{\mu}_{i l}} |_{\lambda^{\mu}_{i l } = 0}$, 
we obtain 
\begin{eqnarray}
  & & 
      U^t_{k,\mu l} 
      = \frac {y^{\mu} (1-r-p)} {2 \sqrt{2 \pi (1-q^t_l)}} 
      \biggl[ e^{\frac {-1} 2 (w^{t+}_{k,\mu l})^2 } - e^{\frac {-1} 2 (w^{t-}_{k,\mu l})^2 } \biggr] \nonumber \\
  & & 
      \qquad \qquad 
      \times \sum_{\tau_l}
      \biggl\{ \tau_l \sgn \biggl[ \sum_{l=1}^K \tau_l \biggr] 
      \nonumber \\ 
  & & 
      \qquad \qquad 
      \times \prod_{l^{\prime} \neq l}^K \biggl[ \frac {1+\tau_{l^{\prime}}} 2 
      - \tau_{l^{\prime}} H (w^{t+}_{k,\mu l^{\prime}}) 
      - \tau_{l^{\prime}} H (w^{t-}_{k,\mu l^{\prime}}) \biggr] \biggr\}, \\
  & & 
      V^t_{k,\mu l} 
      = \frac 1 2 + \frac {y^{\mu}} 2 (r-p) + \frac {y^{\mu}} 2 (1-r-p) \nonumber \\
  & & 
      \qquad \qquad
      \times
      \sum_{\tau_l} \biggl\{ \sgn \biggl[ \sum_{l=1}^K \tau_l \biggr]
      \nonumber \\ 
  & & 
      \qquad \qquad
      \times \prod_{l=1}^K \biggl[ \frac {1+\tau_{l}} 2 
      - \tau_{l} H (w^{t+}_{k,\mu l}) - \tau_{l} H (w^{t-}_{k,\mu l}) \biggr] \biggr\}, \\
  & & 
  \tilde{U}^t_{k,\mu l} 
  = \frac {\left[ w^{t+}_{k,\mu l} e^{\frac {-1} 2 (w^{t+}_{k,\mu l})^2 } 
      + w^{t-}_{k,\mu l} e^{\frac {-1} 2 (w^{t-}_{k,\mu l})^2 } \right] U^t_{k,\mu l}} 
      {\sqrt{1-q^t_l} \left[ e^{\frac {-1} 2 (w^{t+}_{k,\mu l})^2 } - e^{\frac {-1} 2 (w^{t-}_{k,\mu l})^2 } \right]}, \\
  & &
  \tilde{V}^t_{k,\mu l} = - U^t_{k,\mu l}, 
\end{eqnarray}
where 
\begin{eqnarray}
  w^{t+}_{k,\mu l} = \frac {k + \bar{\wedge}^t_{\mu l} - \hat{\wedge}^t_{\mu l} } {\sqrt{1-q^t_l}}, \quad 
  w^{t-}_{k,\mu l} = \frac {k - \bar{\wedge}^t_{\mu l} + \hat{\wedge}^t_{\mu l} } {\sqrt{1-q^t_l}}.
\end{eqnarray}
and $\sum_{\tau_l}$ denotes the sum over all the possible state 
for the dummy binary variables $\{ \tau_l \}$ which can take the value $\pm 1$.

%%%%%%%%%%%%%%%%%%%%%%%%%%%%%%%%%%%%%%%%%%%%%%%%%%%%%%%%%%%%%%%%%%%%%%%%%%%%%%%%%%%%%%%%%%%%%
\subsection{Committee tree with a non-monotonic output unit (CTO)}

\par
In this case it should be noted that optimal performance are obtain 
only for a number of hidden unit $K \to \infty$. 
However we decide to investigate the performance given 
by the scheme even with a finite number of hidden units. 
In the case of the CTO $\mathcal{F}_k$ is given by, 
\begin{eqnarray}
  & &\mathcal{F}_k \left( \left\{ \sqrt{\frac K N} \vec{s}_l \cdot \vec{x}_l^{\mu} \right\} \right) \nonumber \\
  & & = f_k \biggl[ \sqrt{\frac 1 K} \sum_{l=1}^{K} \sgn \biggl( \sqrt{\frac K N} \vec{s}_l \cdot \vec{x}_l^{\mu} \biggr) \biggr] \nonumber \\
  & & \approx f_k \biggl[ \sqrt{\frac 1 K} \sgn \biggl( \lambda^{\mu}_{i l} + \wedge^{\mu}_{i l} \biggr) 
      + \sqrt{\frac 1 K} \sum_{l^{\prime} \neq l}^{K} \sgn \biggl( \wedge^{\mu}_{i l^{\prime}} \biggr) \biggr].
\end{eqnarray}
In the same way as the PTH, we find 
\begin{eqnarray}
  \mathfrak{F}^t_{k,\mu i l} (\lambda^{\mu}_{i l}) 
  & = &
  \int_{-\infty}^{+\infty} \prod_{l=1}^K Dz_l 
  \biggl\{ \frac 1 2 + \frac {y^{\mu}} 2 (1-r-p) 
  \nonumber \\ 
  & &
  \times f_k \biggl[ \sqrt{\frac 1 K} \sgn \biggl( \lambda^{\mu}_{i l} + \bar{\wedge}^{t}_{\mu i l} + z_l \sqrt{1-q^{t}_{\mu i l}} \biggr) \nonumber \\
  & &
  + \sqrt{\frac 1 K} \sum_{l^{\prime} \neq l}^{K} \sgn \biggl( \bar{\wedge}^{t}_{\mu i l^{\prime}} + z_{l^{\prime}} \sqrt{1-q^{t}_{\mu i l^{\prime}}} \biggr) \biggr] 
  \nonumber \\ 
  & &
  + \frac {y^{\mu}} 2 (r-p) \biggr\}, 
\end{eqnarray}
and have 
\begin{eqnarray}
  & & 
  U^t_{k,\mu l} = \frac {y^{\mu} (1-r-p)} {2 \sqrt{2 \pi (1-q^t_l)}} 
  e^{ - \frac 12 (w^{t}_{k,\mu l})^2 } \nonumber \\ 
  & &
  \qquad \qquad
  \times \sum_{\tau_l}
  \left\{ \tau_l f_k \left[ \sum_{l=1}^K \frac{\tau_l} {\sqrt{K}} \right] 
  \prod_{l^{\prime} \neq l}^K H [- \tau_{l^{\prime}} w^{t}_{k,\mu l^{\prime}}] \right\}, \\
  & & 
  V^t_{k,\mu l} = \frac 1 2 + \frac {y^{\mu}} 2 (r-p) + \frac {y^{\mu}} 2 (1-r-p) 
  \nonumber \\ 
  & &
  \qquad \qquad 
  \times
  \sum_{\tau_l} \left\{ f_k \left[ \sum_{l=1}^K \frac{\tau_l} {\sqrt{K}} \right]
  \prod_{l=1}^K H [- \tau_{l} w^{t}_{k,\mu l}] \right\}, \\
  & & 
  \tilde{U}^t_{k,\mu l} = \frac 1 {\sqrt{1-q^t_l}} w^{t}_{k,\mu l} U^t_{k,\mu l}, \\
  & & 
  \tilde{V}^t_{k,\mu l} = - U^t_{k,\mu l},
\end{eqnarray}
where
\begin{eqnarray}
  w^{t}_{k,\mu l} = \frac {\bar{\wedge}^t_{\mu l} - \hat{\wedge}^t_{\mu l} } {\sqrt{1-q^t_l}}.
\end{eqnarray}

%%%%%%%%%%%%%%%%%%%%%%%%%%%%%%%%%%%%%%%%%%%%%%%%%%%%%%%%%%%%%%%%%%%%%%%%%%%%%%%%%%%%%%%%%%%%%
%%  APPENDIX B
\section{Derivation of the BP encoder for lossy compression \label{appendix.B}}

\par
In the lossy compression case, $G_{k,\mu}$ is given by 
\begin{eqnarray}
  & & G_{k,\mu} \biggl( \biggl\{ \sqrt{\frac {K} {N}} \vec{s}_l \cdot \vec{x}^{\mu}_l \biggr\} \biggr)  \nonumber \\
  & & = e^{-\beta} + (1-e^{-\beta}) \theta \biggl[ y^{\mu} \mathcal{F}_k \biggl( \biggl\{ \sqrt{\frac K N} \vec{s}_l \cdot \vec{x}_l^{\mu} \biggr\} \biggr)
  \biggr],  \label{GfuncLC}
\end{eqnarray}
according to the reference \cite{Cousseau2008}. 
The method to derive the set of BP messages is exactly same as in the error correcting cases. 
Thus, the BP equations are given by (\ref{BP:m}), (\ref{BP:mMarg}) and (\ref{BP:mHat}) 
for the standard algorithm and by (\ref{refineBP}) for the more approximated version. 
Only $\mathfrak{F}$, $U$, $\tilde{U}$, $V$ and $\tilde{V}$ change. 
Therefore, in lossy compression case, we first calculate 
$U^t_{k,\mu l}$, 
$V^t_{k,\mu l}$, 
$\tilde{U}^t_{k,\mu l}$ and 
$\tilde{V}^t_{k,\mu l}$ 
for each scheme.

%%%%%%%%%%%%%%%%%%%%%%%%%%%%%%%%%%%%%%%%%%%%%%%%%%%%%%%%%%%%%%%%%%%%%%%%%%%%%%%%%%%%%%%%%%%%%
\subsection{Parity tree with non-monotonic hidden units (PTH)}

\par
In the case of the PTH, using the same method as the error correcting case, one can find $\mathfrak{F}_k$, 
\begin{eqnarray}
  \mathfrak{F}^t_{k,\mu i l} (\lambda^{\mu}_{i l}) 
  & = &
  \int_{-\infty}^{+\infty} \prod_{l=1}^K Dz_l \biggl\{ \frac 12 + \frac {y^{\mu}} 2 \nonumber \\ 
  & &
  \times f_k \biggl( \lambda^{\mu}_{i l} + \bar{\wedge}^{t}_{\mu i l} + z_l \sqrt{1-q^{t}_{\mu i l}} \biggr) \nonumber \\ 
  & & \times \prod_{l^{\prime} \neq l}^{K} f_k \biggl( \bar{\wedge}^{t}_{\mu i l^{\prime}} + z_{l^{\prime}} \sqrt{1-q^{t}_{\mu i l^{\prime}}} \biggr) \biggr\}. 
\end{eqnarray}
In the same way, one can obtain 
\begin{eqnarray}
  & & 
  U^t_{k,\mu l} = \frac {y^{\mu} (1-e^{-\beta})} {\sqrt{2 \pi (1-q^t_l)}} 
  \left[ e^{\frac {-1} 2 (w^{t+}_{k,\mu l})^2 } - e^{\frac {-1} 2 (w^{t-}_{k,\mu l})^2 } \right]
  \nonumber \\ 
  & &
  \qquad \qquad 
  \times \prod_{l^{\prime} \neq l}^K \left[ 1 - 2 H (w^{t+}_{k,\mu l^{\prime}}) -2 H (w^{t-}_{k,\mu l^{\prime}}) \right], \\
  & & 
  V^t_{k,\mu l} = e^{-\beta} + (1-e^{-\beta}) \biggl( \frac 1 2 + \frac {y^{\mu}} 2  \nonumber \\
  & & 
  \qquad \qquad 
  \times \prod_{l=1}^K \left[ 1 - 2 H (w^{t+}_{k,\mu l}) -2 H (w^{t-}_{k,\mu l}) \right] \biggr), \\
  & & 
  \tilde{U}^t_{k,\mu l} = 
  \frac {\left[ w^{t+}_{k,\mu l} e^{\frac {-1} 2 (w^{t+}_{k,\mu l})^2 } + w^{t-}_{k,\mu l} e^{\frac {-1} 2 (w^{t-}_{k,\mu l})^2 } \right] U^t_{k,\mu l}} 
  {\sqrt{1-q^t_l} \left[ e^{\frac {-1} 2 (w^{t+}_{k,\mu l})^2 } - e^{\frac {-1} 2 (w^{t-}_{k,\mu l})^2 } \right]}, \\
  & &
  \tilde{V}^t_{k,\mu l} = - U^t_{k,\mu l}, 
\end{eqnarray}
where
\begin{eqnarray}
  w^{t+}_{k,\mu l} = \frac {k + \bar{\wedge}^t_{\mu l} - \hat{\wedge}^t_{\mu l} } {\sqrt{1-q^t_l}}, \quad 
  w^{t-}_{k,\mu l} = \frac {k - \bar{\wedge}^t_{\mu l} + \hat{\wedge}^t_{\mu l} } {\sqrt{1-q^t_l}}.
\end{eqnarray}

%%%%%%%%%%%%%%%%%%%%%%%%%%%%%%%%%%%%%%%%%%%%%%%%%%%%%%%%%%%%%%%%%%%%%%%%%%%%%%%%%%%%%%%%%%%%%
\subsection{Committee tree with non-monotonic hidden units}

\par
In the case of the CTH, one can find 
\begin{eqnarray}
  \mathfrak{F}^t_{k,\mu i l} (\lambda^{\mu}_{i l}) 
  & = &
  \int_{-\infty}^{+\infty} \prod_{l=1}^K Dz_l \nonumber \\
  & &
  \times \Theta \biggl[ y^{\mu} f_k \biggl( \lambda^{\mu}_{i l} + \bar{\wedge}^{t}_{\mu i l} + z_l \sqrt{1-q^{t}_{\mu i l}} \biggr) \nonumber \\ 
  & &
  + y^{\mu} \sum_{l^{\prime} \neq l}^{K} f_k \biggl( \bar{\wedge}^{t}_{\mu i l^{\prime}} + z_{l^{\prime}} \sqrt{1-q^{t}_{\mu i l^{\prime}}} \biggr) \biggr] ,
\end{eqnarray}
where $\Theta$ denotes the unit step function 
which takes $1$ for $x \geq 0$ and 0 for $x<0$. 
Using the above equation, we have 
\begin{eqnarray}
  & & 
  \hspace*{-8mm}
  U^t_{k,\mu l} = \frac {(1-e^{-\beta})} {\sqrt{2 \pi (1-q^t_l)}} 
  \left[ e^{\frac {-1} 2 (w^{t+}_{k,\mu l})^2 } - e^{\frac {-1} 2 (w^{t-}_{k,\mu l})^2 } \right] \nonumber \\
  & &
  \hspace*{-8mm}
  \qquad \qquad 
  \times \sum_{\tau_l}
  \left\{ \tau_l \Theta \left[ y^{\mu} \sum_{l=1}^K \tau_l \right] 
  \right. \nonumber \\ 
  & & 
  \hspace*{-8mm}
  \qquad \qquad
  \left.
  \times \prod_{l^{\prime} \neq l}^K \left[ \frac {1+\tau_{l^{\prime}}} 2 
  - \tau_{l^{\prime}} H (w^{t+}_{k,\mu l^{\prime}}) 
  - \tau_{l^{\prime}} H (w^{t-}_{k,\mu l^{\prime}}) \right] \right\},
  \\
  & & 
  \hspace*{-8mm}
  V^t_{k,\mu l} = e^{-\beta} + (1-e^{-\beta}) \times
  \sum_{\tau_l} \biggl\{ \Theta \biggl[ y^{\mu} \sum_{l=1}^K \tau_l \biggr]
  \nonumber \\ 
  & & 
  \hspace*{-8mm}
  \qquad \qquad 
  \times
  \prod_{l=1}^K \biggl[ \frac {1+\tau_{l}} 2 - \tau_{l} H (w^{t+}_{k,\mu l}) 
  - \tau_{l} H (w^{t-}_{k,\mu l}) \biggr] \biggr\}, \\
  & &
  \hspace*{-8mm}
  \tilde{U}^t_{k,\mu l} = 
  \frac {\left[ w^{t+}_{k,\mu l} e^{\frac {-1} 2 (w^{t+}_{k,\mu l})^2 } 
  + w^{t-}_{k,\mu l} e^{\frac {-1} 2 (w^{t-}_{k,\mu l})^2 } \right]U^t_{k,\mu l}} 
  {\sqrt{1-q^t_l}\left[ e^{\frac {-1} 2 (w^{t+}_{k,\mu l})^2 } - e^{\frac {-1} 2 (w^{t-}_{k,\mu l})^2 } \right]}, \\
  & & 
  \hspace*{-8mm}
  \tilde{V}^t_{k,\mu l} = - U^t_{k,\mu l},
\end{eqnarray}
where 
\begin{eqnarray}
  w^{t+}_{k,\mu l} = \frac {k + \bar{\wedge}^t_{\mu l} - \hat{\wedge}^t_{\mu l} } {\sqrt{1-q^t_l}}, \quad 
  w^{t-}_{k,\mu l} = \frac {k - \bar{\wedge}^t_{\mu l} + \hat{\wedge}^t_{\mu l} } {\sqrt{1-q^t_l}}.
\end{eqnarray}

%%%%%%%%%%%%%%%%%%%%%%%%%%%%%%%%%%%%%%%%%%%%%%%%%%%%%%%%%%%%%%%%%%%%%%%%%%%%%%%%%%%%%%%%%%%%%
\subsection{Committee tree with a non-monotonic output unit (CTO)}

\par
In this case it should be noted that optimal performance are obtain 
only for  a number of hidden unit $K \to \infty$. 
However we decide to investigate the performance given by the scheme 
even with a finite number of hidden unit. 
We find $\mathfrak{F}_k$ as follows: 
\begin{eqnarray}
  \mathfrak{F}^t_{k,\mu i l} (\lambda^{\mu}_{i l}) 
  & = &
  \int_{-\infty}^{+\infty} \prod_{l=1}^K Dz_l 
  \Theta \bigg(
  y^{\mu} f_k \bigg[ \nonumber \\
  & &
  \sqrt{\frac 1 K} \sgn \bigg( \lambda^{\mu}_{i l} 
  + \bar{\wedge}^{t}_{\mu i l} + z_l \sqrt{1-q^{t}_{\mu i l}} \bigg) 
  \nonumber \\ 
  & &
  + \sqrt{\frac 1 K} \sum_{l^{\prime} \neq l}^{K} \sgn \bigg( \bar{\wedge}^{t}_{\mu i l^{\prime}} 
  + z_{l^{\prime}} \sqrt{1-q^{t}_{\mu i l^{\prime}}} \bigg) \bigg] \bigg) . 
\end{eqnarray}
We then have 
\begin{eqnarray}
  & &
  U^t_{k,\mu l} = \frac {(1-e^{-\beta})} {\sqrt{2 \pi (1-q^t_l)}} e^{- \frac 12 (w^{t}_{k,\mu l})^2 } 
  \nonumber \\ 
  & &
  \qquad \qquad 
  \times \sum_{\tau_l} \bigg\{ \tau_l \Theta \bigg[ y^{\mu} f_k \left( \sum_{l=1}^K \frac{\tau_l} {\sqrt{K}} \right) \bigg] \nonumber \\ 
  & &
  \qquad \qquad 
  \times \prod_{l^{\prime} \neq l}^K H [- \tau_{l^{\prime}} w^{t}_{k,\mu l^{\prime}}] \bigg\}, 
  \\
  & & 
  V^t_{k,\mu l} = e^{-\beta} + (1-e^{-\beta}) 
  \nonumber \\ 
  & &
  \qquad \qquad 
  \times
  \sum_{\tau_l} \bigg\{ \Theta \bigg[ y^{\mu} f_k \left( \sum_{l=1}^K \frac{\tau_l} {\sqrt{K}} \right) \bigg] \nonumber \\
  & &
  \qquad \qquad 
  \times \prod_{l=1}^K H [- \tau_{l} w^{t}_{k,\mu l}] \bigg\},
  \\
  & &
  \tilde{U}^t_{k,\mu l} = \frac 1 {\sqrt{1-q^t_l}} w^{t}_{k,\mu l} U^t_{k,\mu l},
  \\
  & &
  \tilde{V}^t_{k,\mu l} = - U^t_{k,\mu l},
\end{eqnarray}
where 
\begin{eqnarray}
  w^{t}_{k,\mu l} = \frac {\bar{\wedge}^t_{\mu l} - \hat{\wedge}^t_{\mu l} } {\sqrt{1-q^t_l}}. 
\end{eqnarray}

%%%%%%%%%%%%%%%%%%%%%%%%%%%%%%%%%%%%%%%%%%%%%%%%%%%%%%%%%%%%%%%%%%%%%%%%%%%%%%%%%%%%%%%%%%%%%
%%%%%%%%%%%%%%%%%%%%%%%%%%%%%%%%%%%%%%%%%%%%%%%%%%%%%%%%%%%%%%%%%%%%%%%%%%%%%%%%%%%%%%%%%%%%%

\end{document}